
\def\title{Quantum affine symmetry in vertex models}
\def\author{Makoto Idzumi${}^a$, Kenji Iohara ${}^b$, Michio Jimbo${}^b$, \cr
Tetsuji Miwa${}^c$, Toshiki Nakashima${}^d$
and Tetsuji Tokihiro${}^a$\cr
{}\cr
${}^a$
Department of Applied Physics, Faculty of Engineering,\cr
The University of Tokyo, Japan \cr
${}^b$
Department of Mathematics, Faculty of Science,\cr
Kyoto University, Kyoto 606, Japan \cr
{}\cr
${}^c$
Research Institute for Mathematical Sciences,\cr
Kyoto University, Kyoto 606, Japan\cr
{}\cr
${}^d$
Department of Mathematical Science,
Faculty of Engineering Science,\cr
Osaka University, Toyonaka, Osaka 560, Japan \cr
{}\cr}
\par
\def\rhead{Quantum affine symmetry in vertex models}
\def\lhead{Idzumi et al.}

\hsize=5.3truein
\vsize=7.8truein
\baselineskip=10pt
\font\twelvebf=cmbx12
\nopagenumbers
\headline={\ifnum\pageno=1 \hss\sl RIMS\hss \else \hdline\fi}
\def\hdline{\ifodd\pageno\rightheadline \else\leftheadline\fi}
\def\rightheadline{\tenrm\hfil{\it \rhead}\hfil\folio}
\def\leftheadline{\tenrm\folio\hfil{\it \lhead}\hfil}
\voffset=2\baselineskip
\vglue 2cm
%
%
\centerline{\twelvebf
\vbox{
\halign{\hfil # \hfil\cr
\title\crcr}}}
%
%
\bigskip
%
%
\centerline{\tenrm
\vbox{
\halign{\hfil#\hfil\cr
\author\crcr}}}
%
%
\vglue 1cm
%
%
%
\def\address #1
{\vglue 0.5cm
\halign{\quad\it##\hfil\cr
#1\crcr}}
\vskip 1.5cm

\def\C{{\bf C}}

\def\Q{{\bf Q}}
\def\Z{{\bf Z}}

\def\F{{\cal F}}

\def\la{\lambda}
\def\La{\Lambda}

\def\Tau{{\cal T}}

\def\Vh{\widehat{V}}
\def\Vk{V^{(k)}}
\def\V1{V^{(1)}}

\def\ep{\varepsilon}

\def\Hom{\hbox{Hom}}
\def\End{\hbox{End}}
\def\id{\hbox{id}}
\def\sid{{\scriptstyle id}}
\def\tr{\hbox{tr}}

\def\slt{\goth{sl}_2}
\def\slth{\widehat{\goth{sl}}_2\hskip 2pt}
\def\uqg{U_q(\goth{g})}
\def\goto#1{{\buildrel #1 \over \longrightarrow}}
\def\br#1{\langle #1 \rangle}
\def\bra#1{\langle #1 |}
\def\ket#1{|#1\rangle}
\def\brak#1#2{\langle #1|#2\rangle}

\def\vac{|\hbox{vac}\rangle}
\def\dvac{\langle \hbox{vac}|}
\def\wt{\hbox{wt}\,}
\def\Phit{\widetilde{\Phi}}

\def\ft{\tilde{f}}
\def\et{\tilde{e}}
\def\fti{\tilde{f}_i}
\def\eti{\tilde{e}_i}
%
\font\germ=eufm10
\def\goth#1{\hbox{\germ #1}}
%
%
\def\Figure(#1|#2|#3)
{\midinsert
\vskip #2
\hsize 9cm
\raggedright
\noindent
{\bf Figure #1\quad} #3
\endinsert}
%
%
\def\Table #1. \size #2 \caption #3
{\midinsert
\vskip #2
\hsize 7cm
\raggedright
\noindent
{\bf Table #1.} #3
\endinsert}
%
\def\sectiontitle#1\par{\vskip0pt plus.1\vsize\penalty-250
 \vskip0pt plus-.1\vsize\bigskip\vskip\parskip
 \message{#1}\leftline{\bf#1}\nobreak\vglue 5pt}
\def\qed{\hbox{${\vcenter{\vbox{
    \hrule height 0.4pt\hbox{\vrule width 0.4pt height 6pt
    \kern5pt\vrule width 0.4pt}\hrule height 0.4pt}}}$}}
\def\subsec(#1|#2){\medskip\noindent{\it #1}\hskip8pt{\it #2}\quad}
\def\eq#1\endeq
{$$\eqalignno{#1}$$}
\def\leq#1\endeq
{$$\leqalignno{#1}$$}
%
%
\def\qbox#1{\quad\hbox{#1}\quad}
%
%

%
%

%
%

%
%

%
%
\def\Proof{\noindent {\sl Proof.\quad}}

%
%
\def\Example{\smallskip\noindent {\sl Example.\quad}}
%
%
\def\Definition#1.#2{\smallskip\noindent {\sl Definition #1.#2\quad}}
%
%
\def\subsec(#1|#2){\medskip\noindent#1\hskip8pt{\sl #2}\quad}
%
%
%
%
\def\abstract#1\endabstract{
\bigskip
\itemitem{{}}
{\bf Abstract.}
\quad
#1
\bigskip
}
%
%
%
\def\sec(#1){Sect.\hskip2pt#1}

\par
{\narrower\bigskip{\noindent\bf Abstract.\quad}
We study the higher spin anologs of the six vertex model on the basis
of its symmetry under the quantum affine algebra $U_q(\slth)$.
Using the method developed recently for the XXZ spin chain, we
formulate the space of states, transfer matrix, vacuum, creation/annihilation
operators of particles, and local operators, purely in the language
of representation theory.
We find that, regardless of the level of the representation involved,
the particles have spin $1/2$, and that the $n$-particle space
has an RSOS-type structure rather than a simple tensor product
of the $1$-particle space.
This agrees with the picture proposed earlier by Reshetikhin.
\bigskip}
\par

\message{Cross-reference macros, B. Davies, version 6 Aug 1992.}

\catcode`@=11


\newif\if@xrf\@xrffalse   
\def\l@bel #1 #2 #3>>{\expandafter\gdef\csname @@#1#2\endcsname{#3}}
\immediate\newread\xrffile
\def\xrf@n#1#2{\expandafter\expandafter\expandafter
\csname immediate\endcsname\csname #1\endcsname\xrffile#2}
\def\xrf@@n{\if@xrf\relax\else%
  \expandafter\xrf@n{openin}{ = \jobname.xrf}\relax%
  \ifeof\xrffile%
    \message{ no file \jobname.xrf - run again for correct forward references
}%
  \else%
    \expandafter\xrf@n{closein}{}\relax%
    \catcode`@=11 \input\jobname.xrf \catcode`@=12%
  \fi\global\@xrftrue%
  \expandafter\expandafter\csname immediate\endcsname%
  \csname  newwrite\endcsname\xrffile%
  \expandafter\xrf@n{openout}{ = \jobname.xrf}\relax\fi}


\newcount\t@g

\def\order#1{%
  \expandafter\expandafter\csname newcount\endcsname
  \csname t@g#1\endcsname\csname t@g#1\endcsname=0
  \expandafter\expandafter\csname newcount\endcsname
  \csname t@ghd#1\endcsname\csname t@ghd#1\endcsname=0

  \expandafter\def\csname #1\endcsname##1{\xrf@@n\csname n@#1\endcsname##1:>}

  \expandafter\def\csname n@#1\endcsname##1:##2>%
    {\def\n@xt{##1}\ifx\n@xt\empty%
     \expandafter\csname n@@#1\endcsname##1:##2:>
     \else\def\n@xt{##2}\ifx\n@xt\empty%
     \expandafter\csname n@@#1\endcsname\unp@ck##1 >:##2:>\else%
     \expandafter\csname n@@#1\endcsname\unp@ck##1 >:##2>\fi\fi}

  \expandafter\def\csname n@@#1\endcsname##1:##2:>%
    {\edef\t@g{\csname t@g#1\endcsname}\edef\t@@ghd{\csname t@ghd#1\endcsname}%
     \ifnum\t@@ghd=\t@ghd\else\global\t@@ghd=\number\t@ghd\global\t@g=0\fi%
     \ifunc@lled{@#1}{##1}\global\advance\t@g by 1%
       {\def\n@xt{##1}\ifx\n@xt\empty%
       \else\writ@new{#1}{##1}{\pret@g\t@ghead\number\t@g}\expandafter%
       \xdef\csname @#1##1\endcsname{\pret@g\t@ghead\number\t@g}\fi}%
       {\pret@g\t@ghead\number\t@g}%
     \else\def\n@xt{##1}%
       \w@rnmess#1,\n@xt>\csname @#1##1\endcsname%
     \fi##2}%

  \expandafter\def\csname ref#1\endcsname##1{\xrf@@n%
     \expandafter\each@rg\csname #1cite\endcsname{##1}}

  \expandafter\def\csname #1cite\endcsname##1:##2,%
    {\def\n@xt{##2}\ifx\n@xt\empty%
     \csname #1cit@\endcsname##1:##2:,\else%
       \csname #1cit@\endcsname##1:##2,\fi}

  \expandafter\def\csname #1cit@\endcsname##1:##2:,%
    {\def\n@xt{\unp@ck##1 >}\ifunc@lled{@#1}{\n@xt}%
      {\expandafter\ifx\csname @@#1\n@xt\endcsname\relax%
       \und@fmess#1,\n@xt>>>\n@xt<<%
       \else\csname @@#1\n@xt\endcsname##2\fi}%
     \else\csname @#1\n@xt\endcsname##2%
     \fi}}


\def\each@rg#1#2{{\let\thecsname=#1\expandafter\first@rg#2,\end,}}
\def\first@rg#1,{\callr@nge{#1}\apply@rg}
\def\apply@rg#1,{\ifx\end#1\let\n@xt=\relax%
\else,\callr@nge{#1}\let\n@xt=\apply@rg\fi\n@xt}

\def\callr@nge#1{\calldor@nge#1-\end-}
\def\callr@ngeat#1\end-{#1}
\def\calldor@nge#1-#2-{\ifx\end#2\thecsname#1:,%
  \else\thecsname#1:,\hbox{\rm--}\thecsname#2:,\callr@ngeat\fi}


\def\unp@ck#1 #2>{\unp@@k#1@> @>>}
\def\unp@@k#1 #2>>{\ifx#2@\@np@@k#1\else\@np@@k#1@> \unp@@k#2>>\fi}
\def\@np@@k#1#2#3>{\ifx#2@\@@np@@k#1>\else\@@np@@k#1>\@np@@k#2#3>\fi}
\def\@@np@@k#1>{\ifcat#1\alpha\expandafter\@@np@@@k\string#1>\else#1\fi}
\def\@@np@@@k#1#2>{@#2}


\def\writ@new#1#2#3{\xrf@@n\immediate\write\xrffile
  {\noexpand\l@bel #1 #2 {#3}>>}}


\def\ifunc@lled#1#2{\expandafter\ifx\csname #1#2\endcsname\relax}
\def\und@fmess#1#2,#3>{\ifx#1@%
  \message{ ** error - eqn label >>#3<< is undefined ** }\else
  \message{ ** error - #1#2 label >>#3<< is undefined ** }\fi}
\def\w@rnmess#1#2,#3>{\ifx#1@%
  \message{ Warning - duplicate eqn label >>#3<< }\else
  \message{ Warning - duplicate #1#2 label >>#3<< }\fi}


\def\t@ghead{}
\newcount\t@ghd\t@ghd=0
\def\taghead#1{\gdef\t@ghead{#1}\global\advance\t@ghd by 1}


\order{@qn}


\let\eqno@@=\eqno
\def\eqno(#1){\xrf@@n\eqno@@\hbox{{\rm(}$\@qn{#1}${\rm)}}}

\let\leqno@@=\leqno
\def\leqno(#1){\xrf@@n\leqno@@\hbox{{\rm(}$\@qn{#1}${\rm)}}}

\def\refeq#1{\xrf@@n{{\rm(}$\ref@qn{#1}${\rm)}}}


\def\eqalignno#1{\xrf@@n\displ@y \tabskip=\centering
  \halign to\displaywidth{\hfil$\displaystyle{##}$\tabskip=0pt
   &$\displaystyle{{}##}$\hfil\tabskip=\centering
   &\llap{$\eqaln@##$}\tabskip=0pt\crcr
   #1\crcr}}

\def\leqalignno#1{\xrf@@n\displ@y \tabskip=\centering
  \halign to\displaywidth{\hfil$\displaystyle{##}$\tabskip=0pt
   &$\displaystyle{{}##}$\hfil\tabskip=\centering
    &\kern-\displaywidth\rlap{$\eqaln@##$}\tabskip\displaywidth\crcr
   #1\crcr}}

\def\eqaln@#1#2{\relax\ifcat#1(\expandafter\eqno@\else\fi#1#2}
\def\eqno@(#1){\xrf@@n\hbox{{\rm(}$\@qn{#1}${\rm)}}}


\def\n@@me#1#2>{#2}
\def\numberby#1{\xrf@@n
  \ifx\s@ction\undefined\else
  \expandafter\let\csname\s@@ve\endcsname=\s@ction\fi
  \ifx\subs@ction\undefined\else
  \expandafter\let\csname\subs@@ve\endcsname=\subs@ction\fi
  \numb@rby#1,>#1>}
\def\numb@rby#1,#2>#3>{\def\n@xt{#1}\ifx\n@xt\empty\taghead{}\else
  \def\n@xt{#2}\ifx\n@xt\empty\n@by#3>\else\n@@by#3>\fi\fi}
\def\n@by#1>{\ifx\s@cno\undefined\expandafter\expandafter
  \csname newcount\endcsname\csname s@cno\endcsname
  \csname s@cno\endcsname=0\else\s@cno=0\fi
  \xdef\s@@ve{\expandafter\n@@me\string#1>}
  \let\s@ction=#1\def#1{\global\advance\s@cno by 1
  \taghead{\number\s@cno.}\s@ction}}
\def\n@@by#1,#2>{\ifx\s@cno\undefined\expandafter\expandafter
  \csname newcount\endcsname\csname s@cno\endcsname
  \csname s@cno\endcsname=0\else\s@cno=0\fi
  \ifx\subs@cst\undefined\expandafter\expandafter
  \csname newcount\endcsname\csname subs@cst\endcsname
  \csname subs@cst\endcsname=0\else\subs@cst=0\fi
  \ifx\subs@cno\undefined\expandafter\expandafter
  \csname newcount\endcsname\csname subs@cno\endcsname
  \csname subs@cno\endcsname=0\else\subs@cno=0\fi
  \xdef\s@@ve{\expandafter\n@@me\string#1>}
  \let\s@ction=#1\def#1{\global\advance\s@cno by 1
  \global\subs@cno=\subs@cst
  \taghead{\number\s@cno.}\s@ction}
  \xdef\subs@@ve{\expandafter\n@@me\string#2>}
  \let\subs@ction=#2\def#2{\global\advance\subs@cno by 1
  \taghead{\number\s@cno.\number\subs@cno.}\subs@ction}}


\def\numberfrom#1{\ifx\s@cno\undefined\else\n@mberfrom#1,>\fi}
\def\n@mberfrom#1,#2>{\def\n@xt{#2}%
  \ifx\n@xt\empty\n@@f#1>\else\n@@@f#1,#2>\fi}
\def\n@@f#1>{\s@cno=#1\advance\s@cno by -1}
\def\n@@@f#1,#2,>{\s@cno=#1\advance\s@cno by -1%
  \subs@cst=#2\advance\subs@cst by -1}


\def\pret@g{}
\def\prefixby#1{\gdef\pret@g{#1}}



\newcount\r@fcount\r@fcount=0
\newcount\r@fcurr
\newcount\r@fone
\newcount\r@ftwo
\newif\ifc@te\c@tefalse
\newif\ifr@feat

\def\refto#1{{\rm[}\def\s@p{}\refn@te#1>>\refc@te#1>>{\rm]}}

\def\refn@te#1>>{\refn@@te#1,>>}

\def\refn@@te#1,#2>>{\r@fnote{\expandafter\unp@ck\str@pbl#1 >> >}%
   \def\n@xt{#2}\ifx\n@xt\empty\else\refn@@te#2>>\fi}

\def\refc@te#1>>{\r@fcurr=0\r@featfalse\def\s@ve{}%
  {\loop\ifnum\r@fcurr<\r@fcount\advance\r@fcurr by 1\c@tefalse%
   \expandafter\refc@@te\number\r@fcurr>>#1,>>%
   \ifc@te\expandafter\refe@t\number\r@fcurr>>\fi\repeat\s@ve}}

\def\refc@@te#1>>#2,#3>>{\def\n@xt{\expandafter\unp@ck\str@pbl#2 >> >}%
   \expandafter\refc@@@te\csname r@f\n@xt\endcsname>>#1>>%
   \def\n@xt{#3}\ifx\n@xt\empty\else\refc@@te#1>>#3>>\fi}

\def\refc@@@te#1>>#2>>{\ifnum#2=#1\relax\c@tetrue\fi}

\def\refe@t#1>>{\ifr@feat\ifnum\r@fone=\r@ftwo\res@cond#1>>%
   \else\reth@rd#1>>\fi\else\r@feattrue\ref@rst#1>>\fi}

\def\ref@rst#1>>{\r@feattrue\r@fone=#1\r@ftwo=#1%
   \s@p\expandafter\relax\number\r@fone}%

\def\res@cond#1>>{\advance\r@ftwo by 1\def\n@xt{#1}%
   \expandafter\ifnum\n@xt=\number\r@ftwo%
   \edef\s@ve{,\expandafter\relax\number\r@ftwo}\else,\ref@rst#1>>\fi}%

\def\reth@rd#1>>{\advance\r@ftwo by 1\def\n@xt{#1}%
   \expandafter\ifnum\n@xt=\number\r@ftwo%
   \edef\s@ve{--\expandafter\relax\number\r@ftwo}\def\s@p{,}\else%
   \def\s@p{,}\s@ve\def\s@ve{}\ref@rst#1>>\fi}%

\def\r@fnote#1%
  {\ifunc@lled{r@f}{#1}\global\advance\r@fcount by 1%
   \expandafter\xdef\csname r@f#1\endcsname{\number\r@fcount}%
   \expandafter\gdef\csname r@ftext\number\r@fcount\endcsname%
   {\message{ Reference #1 to be supplied }%
   Reference $#1$ to be supplied\par}\fi}

\def\str@pbl#1 #2>>{#1#2}


\def\refis#1 #2\par{\def\n@xt{\unp@ck#1 >}\r@fis\n@xt>>#2>>}
\def\r@fis#1>>#2>>{\ifunc@lled{r@f}{#1}\else
   \expandafter\gdef\csname r@ftext\csname
r@f#1\endcsname\endcsname{#2\par}\fi}


\def\listreferences{\global\r@fcurr=0%
  {\loop\ifnum\r@fcurr<\r@fcount\global\advance\r@fcurr by 1%
   \numr@f\number\r@fcurr>>\csname r@ftext\number\r@fcurr\endcsname>>%
   \repeat}}

\def\numr@f#1>>#2>>{\vbox{\noindent\hang\hangindent=30truept%
   {\hbox to 30truept{\rm[#1]\hfill}}#2}\smallskip\par}


\def\printlabels{\global\@xrftrue\def\s@me##1{$##1$}
  \def\@qn##1{##1}\def\refeq##1{{\rm(}$##1${\rm)}}\refbylabel
  \def\numberby##1{\relax}\def\numberfrom##1{\relax}
  \def\listreferences{\relax}\def\referencefile{\relax}
  \def\order##1{\expandafter\let\csname ##1\endcsname=\s@me
                \expandafter\let\csname ref##1\endcsname=\s@me}}

\def\refbylabel{\def\refto##1{[$##1$]}%
  \def\refis##1 ##2\par{\numr@f$##1$>>##2>>}\def\numr@f##1>>##2>>%
  {\noindent\hang\hangindent=30truept{{\rm[##1]}\ }##2\par}}


\def\beginsection#1\par{\vskip0pt plus.3\vsize\penalty-250
  \vskip0pt plus -.3\vsize\bigskip\vskip\parskip
  \leftline{\bf#1}\nobreak\smallskip\noindent}

\catcode`@=12

\overfullrule=0pt

\order{prop}
\numberby{\beginsection}\numberfrom{1}

\def\s(#1,#2){\sigma^{#1}_{#2}}


\beginsection \S 1. Introduction

The spin-1/2 XXZ model has the Hamiltonian
\eq
&{\cal{H}}=-{1\over2}\sum_{l\in\Z}
\bigl(\s(x,l+1)\s(x,l)+\s(y,l+1)\s(y,l)
+\Delta\s(z,l+1)\s(z,l)\bigr)&(0.1)
\endeq
where
$\s(x,l),\s(y,l),\s(z,l)$
are the Pauli matrices acting on the $l$-th component of the infinite tensor
product
\eq
&\cdots\otimes V\otimes V\otimes V\otimes\cdots,
\qquad V=\C^2.
&(0.3)\cr
&\quad\quad~\scriptstyle l+1\quad\scriptstyle l\quad~\,\scriptstyle l-1
\endeq
Integrable generalizations are known in
the representation theory of the quantum affine algebras
$U_q(\widehat{\goth{g}})$ \refto{S}.
Given an $R$-matrix for $U_q(\widehat{\goth{g}})$,
\eq
&R(z_1/z_2)\in{\rm End}_\C(V_{z_1}\otimes V_{z_2}),\cr
\endeq
the corresponding Hamiltonian is  given by
\eq
&{\cal{H}}=\sum_{l\in\Z}h_{l+1\,l},&(0.2)\cr
\endeq
where $h_{l+1\,l}$ is the following operator $h$
acting on the $(l+1)$-th and the $l$-th components:
\eq
&PR(z_1/z_2)={\rm const}(1+u h +\cdots) \qquad u\rightarrow 0,&(000)\cr
&P:\hbox{ the transposition},\quad z_1/z_2=e^u.\cr
\endeq
If $\goth{g}=\slt$ and $V_z$ is the two dimensional
$U_q\bigl(\slth\bigr)$-module, we have the Hamiltonian \refeq{0.1}
with $\Delta=(q+q^{-1})/2$.
An immediate consequence of \refeq{0.2} is that
the Hamiltonian has the quantum affine symmetry, i.e.,
\eq
&[\Delta^{(\infty)}(g),{\cal{H}}]=0,\cr
\endeq
where $g$ is one of the Chevalley generators of
$U_q(\widehat{\goth{g}})$ and $\Delta^{(\infty)}$
is the infinite coproduct which formally realizes the action of $g$
on \refeq{0.3}.

In \refto{DFJMN} a new method was given for
the diagonalization of \refeq{0.1}
in the anti-ferroelectric regime $-1<q<0$.
It exploits the quantum affine symmetry of \refeq{0.1}
and gives all the eigenvectors of \refeq{0.1} in the thermodynamic limit.
As is well known the Yang-Baxter equation for the $R$ matrix entails that
the row transfer matrices $T(z)$ of the corresponding vertex model
(the six vertex model in the case \refeq{0.1}) are mutually commuting.
The relation \refeq{000} says that \refeq{0.1} is the first in the
hierarchy of commuting Hamiltonians derived from the transfer matrices
\eq
&{\cal H}_n=const. \left(z{d\over dz}\right)^n\log T(z) \bigl|_{z=1}.
\endeq
In this paper we will formulate a similar method
for diagnonalizing the transfer matrices of vertex models associated
with $U_q\bigl(\slth\bigr)$.

Consider level $k$ vertex operators in the sense of \refto{FR}:
\eq
&\Phi_\lambda^{\mu V}(z)~:~
V(\lambda)~\longrightarrow~ V(\mu)\widehat{\otimes} V_z&(0.4)\cr
\endeq
where $V(\lambda)$ and $V(\mu)$ are level $k$
$U_q(\widehat{\goth{g}})$-modules and $V$ is finite-dimensional.
Suppose $V$ has a perfect crystal of level $k$
in the sense of \refto{(KMN)^2}.
Then for any level $k$ dominant integral weight $\lambda$,
there exists a unique level $k$ dominant integral weight
$\mu=\sigma(\la)$
such that the vertex operator $\Phi_\lambda^{\mu V}(z)$ exists,
and the latter is unique up to normalization \refto{DJO}.
For example, if $\goth{g}=\slt$, then the $(k+1)$-dimenaional irreducible
$U_q\bigl(\slth \bigr)$-module
$V^{(k)}$ has a perfect crystal of level $k$.
The corresponding vertex model is the spin-$k/2$ integrable generalization
of the six vertex model introduced in \refto{S}.

The main idea of the diagonalization scheme
is to replace the infinite tensor product \refeq{0.3} by the
level $0$ module
\eq
\F=&\oplus_{\la,\la'}
{\rm Hom}_\C\bigl(V(\la'), V(\la)\bigr)
=\oplus_{\la,\la'}
V(\la)\widehat{\otimes} V(\la')^{*a},\cr
\endeq
where $V(\la')^{*a}$ signifies the dual module defined via
the antipode $a$.
The vertex operator \refeq{0.4}
enables one to disclose the `local component' $V^{(k)}_z$ hidden in $V(\la)$:
\eq
&V(\la)\widehat{\otimes} V(\la')^{*a}
{}~{\buildrel\Phi_\la^{\mu V}(z)\otimes \sid \over\longrightarrow}~
V(\mu) \widehat{\otimes}
V^{(k)}_z\widehat{\otimes} V(\la')^{*a},
\qquad \mu=\sigma(\la).\cr
\endeq
The tensor structure \refeq{0.3} is hidden behind the `doors' $V(\la)$
and $V(\la')^{*a}$, and the vertex operators open/close these doors
to show the tensor components.
We postulate that, in the anti-ferroelectric regime of the model,
the space spanned by finite excitations over
the ground states is given precisely by ${\cal{F}}$.
On the basis of this postulate,
the transfer matrices, the norm of vectors,
the vacuum vectors, the creation and the annihilation operators
and local operators
can all be translated in the language of representation theory,
in much the same way as in \refto{DFJMN}.

The creation operator is interpreted as a vertex operator
$V(\lambda)\rightarrow  V^{(l)}_z \widehat{\otimes} V(\mu)$
(called type II in \refto{DFJMN}).
In contrast to the level $1$ case, for $k>1$
there are various candidates for $l=1,\cdots,k$.
The  correct choice is dictated by the knowledge of the particle structure
of the space of states, i.e. the decomposition of
$V(\la)\widehat{\otimes} V(\la')^{*a}$ into the irreducible pieces.
In the $q=0$ limit, the Hamiltonian becomes diagonal and  the
decomposition can be carried through by the crystal base theory.
Namely, we can identify the particles with the connected components
of the crystal
$B(\la)\widehat{\otimes} B(\la')^*$.
In the case of $\goth{g}=\slt$,
even for higher $k$, we find only spin-$1/2$ particles:
the connected components of
$B(\la)\widehat{\otimes} B(\la')^*$
are isomorphic to the connected components of
Aff$(B^{(1)\otimes n})$
where $B^{(1)}$ is the crystal of the spin-$1/2$ (i.e., the two dimensional)
$U_q\bigl(\slth \bigr)$-module.
This is consistent with the known Bethe Ansatz result \refto{Reshet},
and motivates us to formulate the creation operator
by using the vertex operator
\eq
&\Phi_\la^{V^{(1)}\la_\pm}(z):
V(\la)~\longrightarrow~ V^{(1)}_z\widehat{\otimes} V(\la_\pm),\cr
\endeq
where
$\la=(k-j)\la_0+j\la_1$
and
$\la_\pm=(k-j\mp1)\la_0+(j\pm1)\la_1$.
This choice leads us to the correct computation of the momentum, the
energy and the $S$-matrix of the particles.

The paper is organized as follows.
In \sec(2) we describe the decomposition of the crystal
$B(\la)\widehat{\otimes }B(\la')^*$ in the case of $\goth{g}=\slt$.
In \sec(3) we compute the two point correlation functions for
various types of vertex operators.
In \sec(4),  we
introduce the space of states, row transfer matrix and
creation/annihilation operators. We calculate their commutation relations
and the formula for the energy-momentum using the results of \sec(3).
We also introduce the local operators and derive
$q$-KZ type difference equations for their from factors.
They are a lattice analog of  those found by Smirnov \refto{Smbk}
in the massive integrable field theories.
The text is followed by two appendices.
In appendix 1 we give a brief summary of the theory of
the $q$-KZ equation for the
correlation functions of vertex operators \refto{FR}.
Appendix 2 is a summary of elementary properties of the
basic hypergeometric series used in the text.
%
%
%
%

\def\r{l}
\def\lv{k}
\def\Zl{\{0,1,\ldots,k\}}
%
\def\q{\quad}
\def\qq{\qquad}

\def\uq{U_q(\goth{g})}
\def\Example{\smallskip\noindent {\sl Example.\quad}}

\beginsection \S 2. Decomposition of crystals

\vskip 1 truepc

An integrable spin-$\lv/2$ Hamiltonian $\cal H$ is obtained from
the $R$-matrix, $\check{R}(x/y)=PR(x/y)$
$\in \Hom(V^{(\lv)}_x\otimes V^{(\lv)}_y,V^{(\lv)}_y\otimes V^{(\lv)}_x)$.
Choosing an appropriate base $\{u_{\r}\}$ (the crystal base) of $V^{(\lv)}$,
the Hamiltonian is diagonal in the limit of $q=0$.
Hence pure tensors
$$
\cdots\otimes u_{p(2)} \otimes u_{p(1)} \otimes u_{p(0)} \otimes u_{p(-1)}
\otimes\cdots
\eqno(2.1)
$$
are eigenvectors at least formally.
Let $B(\la)$ be the crystal of
the irreducible level $\lv$ highest weight $U_q(\slth)$-module
$V(\la)$ with highest weight $\la$,
and $B(\mu)^{\ast}$ the crystal of
the irreducible level $-\lv$ lowest weight $U_q(\slth)$-module
$V(\mu)^{\ast a}$ with lowest weight $-\mu$.
We know that the crystal $B(\la)\otimes B(\mu)^{\ast}$
of $V(\la) \otimes V(\mu)^{\ast a}$
is identified with a set of vectors of the form (2.1)
with some boundary condition for $\{p(\r)\}$.
It has a graph structure
with respect to the action of $\eti$ and $\fti$
and decomposes into connected components.
In what follows we shall give a description of these connected components.

\vskip 1 truepc

\subsec(2.1| Paths )  A path is a map $p:\Z \to \Zl$.
We write it as
$$p=(\ldots,p(2),p(1),p(0), p(-1),\ldots).$$
For $m\in\Zl$ there is a path defined by
$$
\matrix{
p(\r)   = & m \hfill & \q \hbox{for }\r\equiv 0 \bmod 2,\cr
\hfill = & \lv-m      & \q \hbox{for }\r\equiv 1 \bmod 2.
}
$$
We denote it by ${\bar p}_m$.
Here $\la_m=(\lv-m)\La_0+m\La_1\ (m=0,1,\ldots,\lv)$ are
the level $\lv$ dominant integral weights.
A path is called a $(\la_m,\la_{m'})$-path ($m,m'\in\Zl$)
if it is subject to the boundary condition
$$
\matrix{
p(\r)   = & {\bar p}_m (\r)\hfill & \q \hbox{for }\r\gg 0,\cr
\hfill = & {\bar p}_{m'} (\r)    & \q \hbox{for }\r\ll 0.
}
$$
The reason why we use $\la_m$'s to specify the boundary conditions is
explained at the end of this subsection.
We denote a set of $(\la_m,\la_{m'})$-paths by ${\cal P}^{(m,m')}$, and write
$$
{\cal P}=\bigsqcup_{(m,m')}{\cal P}^{(m,m')}.
$$
In the following, we consider only paths belonging to ${\cal P}$
and identify a path $p$ with the tensor (2.1).

Let $B^{(k)}=\{u_{\r}\}$ be the crystal base
of the $(\lv+1)$-dimensional (or spin $\lv/2$)
irreducible representation $V^{(\lv)}$ of $U_q(\slth)$
(\refto{Ka}; see also \refto{(KMN)^2}).
The action of the operators $\eti,\fti$($i=0,1$) is given by
$$
\matrix{
\ft_1 u_{\r}   = & u_{\r+1}   & \q \hbox{for }0\le  \r<\lv, \cr
\hfill      = & 0 \hfill  & \q \hbox{for }\r=\lv;\hfill \cr
\ft_0 u_{\r}   = & u_{\r-1}   & \q \hbox{for }0<\r\le  \lv, \cr
\hfill      = & 0 \hfill  & \q \hbox{for }\r=0;\hfill
}
$$
and $\et_1=\ft_0$, $\et_0=\ft_1$.

Let $B=\{v_+,v_-\}$ be the $U_q\bigl(\slt \bigr)$ crystal
such that
$\ft v_{+}= v_{-}$, $\ft v_{-}= 0$,
$\et v_{+}= 0$, and $\et v_{-}= v_{+}$.
When the color $i=0,1$ is fixed,
it is convenient to embed $B^{(k)}$ (considered
as a $U_q\bigl(\slt \bigr)$ crystal via $\ft=\fti,\et=\eti$)
into the $\lv$-fold tensor product $B^{\otimes \lv}$:
For $i=1$(resp. $0$), $u_{\r}$ is replaced by
$$
\underbrace{v_{-}\otimes\cdots\otimes v_{-}}_{\r(\hbox{resp. }\lv-\r)}
           \otimes
           \underbrace{v_{+}\otimes\cdots\otimes v_{+}}_{\lv-\r(\hbox{resp.
}\r)}.
\eqno(2.2)
$$

Now we describe the rules for the action of $\eti,\fti$ on ${\cal P}$,
which is obtained from the knowledge of the action on tensor products
(later in \sec(2.2) we shall give much simpler rules).
For $p\in {\cal P}$,  $\fti p \in {\cal P}\cup \{0\}$
is determined as follows: \hfill\break
(i) Using (2.2) replace each component of the path by $-$'s and $+$'s:
for $i=1$(resp. $0$),
$$
p(\r) \mapsto
(  \underbrace{- \q - \q \cdots\cdots \q -}_{p(\r)(\hbox{resp. }\lv-p(\r))}
\q \underbrace{+ \q + \q \cdots\cdots \q +}_{\lv-p(\r)(\hbox{resp. }p(\r))}).
$$
Then we get an infinite $\pm$ sequence, denoted by ${\cal C}(p)$.
\hfill\break
(ii) Remove `singlet pairs' $+-$ successively from ${\cal C}(p)$,
and then ${\cal C}(p)$ is reduced to a finite sequence
of the form $(- \cdots - + \cdots +)$.
We denote it by $\bar{\cal C}(p)$.
\hfill\break
(iii) The action of $\fti$ on $\bar{\cal C}(p)$
changes the left most $+$ to $-$.
If there is no $+$ in $\bar{\cal C}(p)$, then $\fti \bar{\cal C}(p)=0$.
\hfill\break
(iv) The action of $\fti$ on ${\cal C}(p)$ changes the $+$,
which corresponds to the one flipped in (iii), to $-$.
If $\fti \bar{\cal C}(p)=0$ then $\fti {\cal C}(p)=0$.
\hfill\break
(v) $\fti p$ is obtained from $\fti {\cal C}(p)$
by reversing the replacement of (i).
If $\fti {\cal C}(p)=0$ then $\fti p=0$.
\hfill\break

\Example
Let $k=2$ and
$p=(\ldots,0,2,1,2,0,2,\ldots)$.
For $i=1$,
$$
\matrix{
\hfill{\cal C}(p)=&
\cdots & + & + & - & - & - & + & - & - & + & + & - & - & \cdots \cr
     & &   &   &   &   & \downarrow &  &  &  &  &  &  &  &      \cr
\hfill\bar{\cal C}(p)=&
       &   &   &   &   & - &   &   & - &   &   &   &   &        \cr
     & &   &   &   &   & \downarrow & \ft_1 &  &  &  &  &  &  & \cr
\hfill\ft_1\bar{\cal C}(p)=&
       &   &   &   &   & 0 &   &   &   &   &   &   &   &        \cr
     & &   &   &   &   & \downarrow &  &  &  &  &  &  &  &      \cr
\hfill\ft_1{\cal C}(p)=&
       &   &   &   &   & 0 &   &   &   &   &   &   &   &        \cr
}
$$
i.e., $\ft_1 p = 0$. \hfill\break
For $i=0$,
$$
\matrix{
\hfill{\cal C}(p)=&
\cdots & - & - & + & + & - & + & + & + & - & - & + & + & \cdots \cr
     & &   &   &   &   & \downarrow &  &  &  &  &  &  &  &      \cr
\hfill\bar{\cal C}(p)=&
       &   &   & + &   &   & + &   &   &   &   &   &   &        \cr
     & &   &   &   &   & \downarrow & \ft_0 &  &  &  &  &  &  & \cr
\hfill\ft_0\bar{\cal C}(p)=&
       &   &   & - &   &   & + &   &   &   &   &   &   &        \cr
     & &   &   &   &   & \downarrow &  &  &  &  &  &  &  &      \cr
\hfill\ft_0{\cal C}(p)=&
\cdots & - & - & - & + & - & + & + & + & - & - & + & + & \cdots \cr
}
$$
i.e., $\ft_0 p = (\ldots,0,1,1,2,0,2,\ldots)$.

\vskip 1 truepc

The action of $\eti$ on $p\in {\cal P}$ is such that
if $\{ p'\in{\cal P} \mid \fti p'=p \}$ is empty then $\eti p=0$;
otherwise $\eti p=p'$ where $p'$ is the unique one satisfying $\fti p'=p$.
Note that only one component, say $p(\r)$, of the path $p$ is affected
by an application of one operator: $p(\r)\mapsto p(\r)\pm 1$
where $+$  is chosen for $\ft_1,\et_0$
and $-$ for $\ft_0, \et_1$.

Besides the action of operators
we define the weight of a path.
For a path $p\in {\cal P}^{(m,m')}$
$$
\eqalignno{
\hbox{wt}(p) & =(m-m')(\La_1-\La_0)+s(p)\alpha_1-h(p)\delta \cr
             & =\Big({1\over 2}m-{1\over 2}m'+s(p)\Big)\alpha_1-h(p)\delta.
}
$$
We note $2(\La_1-\La_0)=\alpha_1$ and $\delta=\alpha_0+\alpha_1$.
Here the integer-valued function $h:{\cal P}^{(m,m')}\to{\bf Z}$ is defined
by
$$
h(p) = \sum_{\r\in{\bf Z}} \r \Big( H(p(\r+1),p(\r))-
                H(\bar{p}_{m,m'}(\r+1),\bar{p}_{m,m'}(\r)) \Big),
$$
where $\bar{p}_{m,m'}(\r)=\bar{p}_{m}(\r)$ for $\r>0$,
$=\bar{p}_{m'}(\r)$ for $\r\le 0$,
and
$$
\matrix{
H(j,j') = & -j'  \hfill & \q \hbox{for }j+j' \le \lv \hfill \cr
\hfill = & j-\lv        & \q \hbox{for }j+j' \ge \lv,\hfill
}
$$
$j,j'=0,1,\ldots,\lv$
($H$ is called energy function);
the `spin-deviation' of the path
$s:{\cal P}^{(m,m')}\to{\bf Z}$ is defined by
$$
s(p) = \sum_{\r\in{\bf Z}} (\bar{p}_{m,m'}(\r)-p(\r)).
$$

Equipped with the action of operators and the weight,
the set of $(\la_m,\la_{m'})$-paths ${\cal P}^{(m,m')}$ is identified with
the crystal $B(\la_m) \otimes B(\la_{m'})^{\ast}$ of
$V(\la_m) \otimes V(\la_{m'})^{\ast a}$.
In particular, $\bar{p}_m$ is the unique singlet in ${\cal P}^{(m,m)}$, which
we call ground state path.
Drawing an arrow $p \buildrel i \over \rightarrow p'$ when $p'=\fti p$,
we equip ${\cal P}$ with a structure of colored oriented graph.
It decomposes into a number of (in fact infinite) connected parts.

\vskip 1 truepc

\subsec(2.2| Domain walls)  An alternative description of paths is obtained
when one observes the deviation from the ground state path $\bar{p}_m$.

Fix a path $p$ for a while.
A segment $(p(\r_1), p(\r_1 -1), \ldots, p(\r_2))$, $\r_1 \geq \r_2$, is
called $\la_m$-domain ($m=0,1,\ldots,\lv$) if
$$
\matrix{
p(\r)   = & {\bar p}_m(\r)\hfill & \q \hbox{for }\r_1 \ge \r \ge \r_2,\hfill
\cr
p(\r)\not=& {\bar p}_m(\r)\hfill & \q \hbox{for }\r=\r_1+1
\hbox{ and } \r=\r_2-1.\hfill
}
$$
The integer $\r_1-\r_2+1$ and
the weight $\la_m$ are called length and type of the domain,
respectively.
A succesive pair $\bigl(p(l+1),p(l)\bigr)$ belongs to the same domain
if and only if $p(l+1)+p(l)=k$.
A path is partitioned into several domains.
The boundary between adjacent domains is called (domain) wall.
If it occurs at $(p(\r+1),p(\r))$, we say the position of the wall is $\r$.

A domain wall is  an elementary wall
if the types of adjacent domains are successive;
i.e., if a $\la_m$-domain and a $\la_{m+1}$-domain are adjacent
then the wall between them is elementary.
Note that if a wall at $\r$ is elementary, $p(\r+1)+p(\r)=\lv+1$ or $\lv-1$.
Accordingly we call the wall is of type 0 (0-wall) or of type 1 (1-wall)
respectively
(As we shall see the $i$-wall may change its position by the action of $\fti$
but never by $\ft_{1-i}$).
A wall which is not elementary is thought to be a composition of
elementary walls:
Let $\la_m$ and $\la_{m'} (m'>m+1)$ be the types of domains
left and right to the wall at $\r$.
We consider that, at the position of the  wall,
there are $m'-m-1$ domains of length zero
(types of which are $\la_{m+1}, \la_{m+2}, \ldots, \la_{m'-1}$
from left to right)
and $m'-m$ elementary walls between them
(types of which are all 0 (resp. 1)
if $p(\r+1)+p(\r)=\lv+(m'-m)$ (resp. $=\lv-(m'-m)$)).
In the case $m>m'+1$ things are similarly defined.

Let $n(p)$ be the number of elementary walls in a path $p$.
Let $(a_0, a_1, \ldots, a_{n(p)})$ be a sequence of types of domains
of the path, counted from left to right,
where  $a_0$ (resp. $a_{n(p)}$) is
the type of the left (resp. right) half infinite domain.
Domains that have zero length are included in the list.
Noting that by definition $a_j-a_{j+1} = \pm(\La_0-\La_1)$,
$j=0,1,\ldots,n(p)-1$,
we can consider that the list $(a_0, a_1, \ldots, a_{n(p)})$ defines
a state (or `path') of length $n(p)$ of the $(k+1)$-state ABF model.
\hfill\break

\Example
Several paths in the case $\lv=2$ (spin 1) are shown.
Elementary walls are denoted by $|$~.
Types of domains $(a_0, a_1, \ldots, a_{n(p)})$ are also given.
$$
\matrix{
\ldots& p(4) & & p(3) & & p(2) & & p(1) & & p(0) & & p(-1)&\ldots
                                                 & \q\hbox{types of domains}\cr
\ldots& 0 & & 2 & & 0 & & 2 &|& 1 & & 1  &\ldots & \q(\la_0,\la_1) \cr
\ldots& 1 & & 1 & & 1 & & 1 &|& 0 & & 2  &\ldots & \q(\la_1,\la_0) \cr
\ldots& 0 & & 2 & & 0 &|& 1 & & 1 & & 1  &\ldots & \q(\la_0,\la_1) \cr
\ldots& 0 & & 2 &|& 1 & & 1 &|& 2 & & 0  &\ldots & \q(\la_0,\la_1,\la_2) \cr
\ldots& 0 & & 2 & & 0 & & 2 &||& 2 & & 0 &\ldots & \q(\la_0,\la_1,\la_2) \cr
}
$$
The last one shows an example of a composite wall.

\vskip 1 truepc

In the language of domain walls,
the action of $\fti, \eti$ can be described most simply.
The rules are obtained if we notice the following:
after the reduction steps (i) and (ii) in \sec(2.1),
a 0-wall is considered to be $v_{-}$
and a 1-wall to be $v_{+}$ with respect to color $i=1$.
With respect to color $i=0$, signs $\pm$ are interchanged.
\hfill\break

\Example
In the case $\lv=2$ (spin 1)
there are four elementary walls $(2,1)$, $(1,2)$, $(0,1)$, and $(1,0)$,
which we can regard as $v_{\pm}$.
Consider the wall (2,1).
For $i=1$, a neighborhood of the wall $(\ldots,0,2,0,2,1,1,1,1,\ldots)$,
is interpreted into
$(\cdots ++--++---+-+-+-+ \cdots)$;
which is reduced to $-$ after successive removement of singlet pairs
(rule (ii)).
Similarly the rest three walls are reduced to $\pm$.
We express these reductions by
$$
\matrix{
21         & 12         & 01         & 10         \cr
---+       & -+--       & ++-+       & -+++       \cr
\downarrow & \downarrow & \downarrow & \downarrow \cr
\q\q-\q    & \q\q\q-    & +\q\q\q    & \q+\q\q    \cr
}
$$
In the case $\lv=3$ (spin 3/2), we have six elementary walls:
for $i=1$,
$$
\matrix{
31         & 22         & 13        \cr
----++     & --+--+     & -++---    \cr
\downarrow & \downarrow & \downarrow\cr
\qq\q-\qq  & \qq\qq-\q  & \qq\qq\q- \cr
  02         & 11         & 20         \cr
  +++--+     & -++-++     & --++++     \cr
  \downarrow & \downarrow & \downarrow \cr
  +\q\qq\qq  & \q+\qq\qq  & \qq+\q\qq  \cr
}
$$

\vskip 1 truepc

Rules for the action of $\ft_1$ (resp. $\ft_0$) are as follows: \hfill\break
(i) Let $n=n(p)$, and let
$(i_1, \ldots, i_n)$ be a sequence of wall-types of $p$
(from left to right).
Replacing 1 by $+$(resp. $-$) and 0 by $-$(resp. $+$) in the sequence,
we get a $\pm$ sequence of length $n$, denoted by ${\cal C}'(p)$.
\hfill\break
(ii) Removing singlet pairs $+-$ successively from ${\cal C}'(p)$,
we get a sequence of the form $(- \cdots - + \cdots +)$,
denoted by $\bar{\cal C}'(p)$.
\hfill\break
(iii) The action of $\fti$ on $\bar{\cal C}'(p)$
changes the left most $+$ to $-$.
If there is no $+$ in $\bar{\cal C}'(p)$, then $\fti \bar{\cal C}'(p)=0$.
\hfill\break
(iv) The action of $\fti$ on ${\cal C}'(p)$ changes the $+$,
which corresponds to the one flipped in (iii), to $-$.
If $\fti \bar{\cal C}'(p)=0$ then $\fti {\cal C}'(p)=0$.
\hfill\break
(v) $\fti p$ is obtained from $\fti {\cal C}'(p)$:
first, by reversing the replacement of (i)
we get a sequence of types of walls;
second, we reconstruct the path $\fti p$ from the
sequence of wall-types.
If $\fti {\cal C}'(p)=0$, we get $\fti p=0$.
\hfill\break

Note that one of the walls shifts to the left
and changes its type ($0 \leftrightarrow 1$) in the action.
The number of walls $n(p)$ and the list of domain-types
$(a_0, a_1, \ldots, a_{n(p)})$ are invariant under the action.

We have a natural decomposition
$$
\eqalignno{
{\cal P}          & = \bigsqcup_{(m,m')}{\cal P}^{(m,m')},\cr
{\cal P}^{(m,m')} & = \bigsqcup_{n=0}^{\infty}{\cal P}^{(m,m')}(n),
}
$$
where ${\cal P}^{(m,m')}(n)$ signifies the set of $(\la_m,\la_{m'})$-paths
with $n$ elementary walls.
A set ${\cal P}^{(m,m')}(0)$ contains only one element $\bar{p}_m$,
the ground state path, if $m=m'$ (otherwise it is empty).
${\cal P}^{(m,m')}(n), n\ge 1$, is an infinite set (or empty).
Define an equivalence relation in a set of paths by
$$
p \sim p' \q \iff \q p'=x_1\cdots x_r p \q
\hbox{ for some }x_j \hbox{'s} \in\{\eti,\fti\mid i=0,1\}.
$$
In other words, $p\sim p'$ if and only if
they belong to the same connected part in ${\cal P}$.
We write $\widetilde{\cal P}={\cal P}/\sim$, and denote its element by $[p]$.
Then
$\widetilde{\cal P}^{(m,m')}(1)$ contains only one element (or is empty), and
$\widetilde{\cal P}^{(m,m')}(n)$ $(n \ge 2)$ contains
infinite number of elements (or is empty).

\vskip 1 truepc

\subsec(2.3| Classification of paths)
It appears that an elementary wall behaves like a local spin-1/2 state.
In fact it is true in the following sense.
Let Aff$(B^{(1)\otimes n})$ be the affinization of $B^{(1)\otimes n}$
(\refto{(KMN)^2}).
Aff$(B^{(1)\otimes n})$ consists of finitely many
connected parts. They are all isomorphic as crystals
without the weight structure.
For convenience, set $v_i=u_{1-i}\in B^{(1)}$.
There is an isomorphism between a connected part $[p]$ in ${\cal P}$
and one of the connected parts of Aff$(B^{(1)\otimes n})$
where $n=n(p)$, induced from the map
$$\iota:[p]\rightarrow B^{(1)\otimes n},\
p'\mapsto
v_{i_1}\otimes \cdots \otimes v_{i_n},$$
where
$(i_1, \ldots, i_n)$
is the sequence of wall-types of $p'\in[p]$.
Physically speaking, there are only spin-1/2 particles (magnons),
but no higher-spin magnons. This agrees with the result of Reshetikhin
\refto{Reshet}.

For a later use we define the `depth' of a path
$\nu(p)$ ($\in{\bf Z}$) by
$$
\nu(p)=-[ \langle \wt(p), d \rangle /n(p) ]
$$
where $\langle \q, \q \rangle: P\times P^{\ast} \to {\bf Z}$,
$\langle \delta, d \rangle =1$, $\langle \alpha_1, d \rangle =0$,
and $[x]$ denotes the integer part of $x$.
Then $\{ ((i_1,\ldots,i_n),\nu) \}\leftrightarrow [p]$ is one to one.

Now we proceed to the classification of connected parts.
We shall parametrize the elements in $\widetilde{\cal P}^{(m,m')}(n)$.
To do this it is necessary to define parameters
related to `lengths of domains'.

Let $\bar{p}$ be a representative of $[p]\in\widetilde{\cal P}$ such that
its sequence of wall-types is $(0,\ldots,0)$.
Let $n=n(\bar{p})$, and let $(a_0, a_1, \ldots, a_{n})$ be
the sequence of domain-types of $\bar{p}$.
For  the $j$-th domain ($j=1,2,\ldots,n-1$)
of $\bar{p}$ (which is an $a_j$-domain)
we define a non-negative integer $m_j \in {\bf Z}_{\ge 0}$ by
$$
\eqalignno{
  2m_j  & = \hbox{ length of domain,}\q \hbox{if }a_{j-1}-a_j=a_j-a_{j+1};\cr
2m_j+1  & = \hbox{ length of domain,}\q \hbox{otherwise.}
}
$$
Then we have
$$
\eqalignno{
\widetilde{\cal P}^{(m,m')}(n)
   & = \bigsqcup_{a_1, \ldots, a_{n-1}}
        \widetilde{\cal P}(n;a_0, a_1, \ldots, a_{n}),\cr
\widetilde{\cal P}(n;a_0, a_1, \ldots, a_{n})
   & \simeq \{(m_1,\ldots,m_{n-1})
        \mid m_j \in {\bf Z}_{\ge 0},  j=1,\ldots,n-1 \}.
}
$$
In the first equation the summation is over all
$a_j\in\{\la_0,\la_1,\ldots,\la_{\lv}\}$, $j=1,\ldots,n-1$,
with restriction $a_j-a_{j+1} = \pm(\La_0-\La_1)$, $j=0,1,\ldots,n-1$,
and $a_0, a_n$ are fixed to $\la_m,\la_{m'}$, respectively.
This completes the classification.

\Example
We give below the
types of domains $(a_0, a_1, \ldots, a_{n(p)})$
and the `length' of domains
$(m_1, \ldots, m_{n(p)-1})$
for several paths in the case $\lv=3$ (spin 3/2).
The asterisk shows the position of $p(0)$.
$$
\matrix{
& \hskip-10pt\ast&&\q\hbox{types of domains} & \q\hbox{`length'} \cr
\ldots&0\ 3|1\ 2\ 1\ 2|2\ 1\ 2\ 1&\ldots & \q(\la_0,\la_1,\la_2) & \q(2) \cr
\ldots&0\ 3|1\ 2\ 1\ 2\ 1|3\ 0\ 3&\ldots & \q(\la_0,\la_1,\la_0) & \q(2) \cr
\ldots&2\ 1\ 2|2\ 1\ 2\ 1\ 2|2\ 1&\ldots & \q(\la_2,\la_1,\la_2) & \q(2) \cr
\ldots&0\ 3|1\ 2\ 1\ 2|2\ 1|3\ 0&\ldots & \q(\la_0,\la_1,\la_2,\la_3) & \q(2,1)
\cr
}
$$

\vskip 1 truepc

Let us summarize the classification of the paths in $\cal{P}$.
The paths are parametrized uniquely by
the number of walls in path $n\in {\bf Z}_{\ge 0}$;
sequence of types of walls $(i_1, \ldots, i_n)$, $i_j=0,1$;
depth parameter  $\nu\in {\bf Z}$;
sequence of types of domains $(a_0, a_1, \ldots, a_{n})$,
$a_j\in\{\la_0,\la_1,\ldots,\la_{\lv}\}$ with $a_j-a_{j+1} =
\pm(\La_0-\La_1)$,
$j=0,1,\ldots,n-1$;
and `length-of-domain' parameters $(m_1,\ldots,m_{n-1})$,
$m_j\in {\bf Z}_{\ge 0}$.
The boundary condition fixes the values of $a_0$ and $a_n$.

\def\qbox#1{\quad\hbox{#1}}

\def\VO(#1,#2){{\tilde \Phi}^{#1}_{#2}}
\def\VVO(#1,#2){{\Phi}^{#1}_{#2}}
\def\h(#1){|u_{#1}\rangle}

\def\lv{\langle{\rm vac}|}

\def\fun(#1,#2,#3,#4,#5,#6){\bra{#1}\Phi^{#2}_{#3}(#4)
                            \Phi^{#5}_{\la}(#6)\ket{\la}}

\def\q{\quad}
\def\qq{\qquad}
\def\pair(#1,#2){{\langle{#1},{#2}\rangle}}
\def\al{\alpha}
\def\ot{\otimes}
\def\V(#1,#2){V^{(#1)}(#2)}
\def\ovl{\overline}
\def\del{\delta}
\def\Del{\Delta}
\def\hg(#1,#2;#3;#4){{}_2\phi_1\left({{{#1}\,\,\,{#2}}\atop{{#3}}};p,
                     {#4}\right)}
\def\hhg(#1,#2;#3;#4){\phi\left({{{#1}\,\,\,{#2}}\atop{{#3}}};
                     {#4}\right)}
\def\wgt(#1,#2){(#1)\La_0+(#2)\La_1}

\def\qbinom#1#2{{#1}\atopwithdelims[]{#2}}
\def\Vk{V^{(k)}}
\def\V1{V^{(1)}}
\def\Vs1{V^{(1)*a^{-1}}}
\def\comp(#1,#2,#3,#4,#5){\Phi^{#1}_{#2}(#3)\Phi^{#4}_{\la}(#5)}

\def\s{\sigma}

\def\pn(#1,#2,#3){{{#1}{{#2}\over{#3}}}}
\def\pp(#1,#2,#3){p^{#1{{#2}\over{#3}}}}
\def\qp(#1,#2,#3){q^{#1{{#2}\over{#3}}}}
\def\zp(#1,#2,#3){z^{#1{{#2}\over{#3}}}}
\def\inp(#1,#2){(#1;#2)_{\infty}}
\def\xx(#1,#2,#3,#4,#5){{\xi({1\over{#5}};#1,#2)\over\xi(#5;#3,#4)}}
\def\qqxx(#1,#2,#3,#4,#5){{\xi({{q^2}\over{#5}};#1,#2)
                            \over\xi({q^{-2}#5};#3,#4)}}
\def\Boltz(#1,#2,#3,#4,#5){W\left(\matrix{{#1}&{#2}\cr
                                 {#3}&{#4}\cr}\biggl|\,{#5}\right)}
\def\BBoltz(#1,#2,#3,#4,#5){\widehat W \left(\matrix{{#1}&{#2}\cr
                                 {#3}&{#4}\cr}\biggl|\,{#5}\right)}

\beginsection \S 3. Two point functions of the vertex operators for $\slth$

\subsec(3.1| Notations) We shall follow the notations of \refto{DFJMN}
unless otherwise stated.
The main change is that here we normalize the invariant bilinear form on
$P=\Z\La_0\oplus\Z\La_1\oplus\Z\delta$ by $(\al_i,\al_i)=2$,
as opposed to the convention $(\al_i,\al_i)=1$ in \refto{DFJMN}.
We have $(\La_i,\La_j)=\delta_{i,1}\delta_{1,j}/2$,
$(\La_i,\delta)=1$ and $(\delta,\delta)=0$.
We identify $P^*=\Z h_0\oplus\Z h_1\oplus \Z d$ with
a subset of $P$ via $(~,~)$, so that
$\al_i=h_i$ and $\rho=\La_0+\La_1=h_1/2+2d$.
We set $P_+=\Z_{\ge 0}\La_0 \oplus \Z_{\ge 0}\La_1$,
$P^0_k=\{ \la\in P_+\mid \br{\la,c}=k \}$.
For $\la\in \Z\La_0\oplus \Z\La_1$
we define $\sigma(\la)$, $\ovl{\la}$ by
\eq
&\sigma(m\La_0+n\La_1)=n\La_0+m\La_1, \quad
\ovl{m\La_0+n\La_1}=n(\La_1-\La_0). &(sigma)
\endeq

In the rest of this paper we denote by $F$ the base field $\Q(q)$.
The quantized affine algebra $U$ is the $F$-algebra generated by $e_i$,
$f_i$ $(i=0,1)$ and $q^h$ ($h\in P^*$) with the defining relations
as in \refto{Ka}, e.g.
$[e_i,f_j]=\delta_{i,j}(t_i-t^{-1}_i)/(q-q^{-1})$ ($t_i=q^{h_i}$).
We take the coproduct
\eq
&\Delta(e_i)=e_i\ot 1+t_i\ot e_i,\q \Delta(f_i)=f_i\ot t_i^{-1}+1\ot f_i,\cr
&\Delta(q^h)=q^h\ot q^h \q(h\in P^*) &(coprod)
\endeq
and the antipode
\eq
a(e_i)&=-t_i^{-1}e_i,\q a(f_i)=-f_it_i,\q a(q^h)=q^{-h}
\q(h\in P^*). &(antipode)
\endeq
$U'$ will denote the subalgebra generated by $e_i$, $f_i$
and $t_i$ $(i=0,1)$.

Let $M$ be the left $U$-module with weight space decomposition
$\oplus_\nu M_\nu$.
By $M^{*a^{\pm 1}}$ we mean
the restricted dual space $M^*=\oplus_\nu M_\nu^*$
endowed with the left module structure via $a^{\pm 1}$:
$$
\pair(xu,v)=\pair(u,{a^{\pm 1}(x)v}) \qbox{for }x\in U,\,u\in M^*,\,v\in M.
$$

The irreducible left (resp. right) highest weight module with highest weight
$\la\in P^0_k$ will be denoted by $V(\la)$ (resp. $V^r(\la)$).
We fix a highest weight vector $\ket{\la}\in V(\la)$
(resp. $\bra{\la} \in V^r(\la)$) once for all.
There is a unique non-degenerate symmetric bilinear
pairing $V^r(\la)\times V(\la)\rightarrow F$ such that
$$
\bra{\la}\la\rangle=1,\q\bra{ux}u'\rangle=\bra{u}xu'\rangle\qbox{for any }
\bra{u}\in V^r(\la),\, \ket{u'}\in V(\la).
$$

\subsec(3.2| $R$-matrices)
For $l\in \Z_{\ge 0}$
let $V^{(l)}$ denote the $(l+1)$-dimensional $U'$-module
with basis $\{v^{(l)}_j\}_{0\le j\le l}$ given by
\eq
&e_1v^{(l)}_j=[j]v^{(l)}_{j-1},\q f_1v^{(l)}_j=[l-j]v^{(l)}_{j+1}, \q
 t_1v^{(l)}_j=q^{l-2j}v^{(l)}_j,\cr
&e_0=f_1, \q f_0=e_1,\q t_0=t_1^{-1}\qbox{on }{V^{(l)}}. &(spinl/2) \cr
\endeq
Here $v^{(l)}_j$ with $j<0$ or $j>l$ is understood to be $0$.
The $\{v^{(l)}_j\}$ constitute an upper global base of $V^{(l)}$ \refto{Ka2}.
In the case $l=1$ we also write $v^{(1)}_0=v_+$ and $v^{(1)}_1=v_-$.

We equip $V_z^{(l)}=V^{(l)}\ot F[z,z^{-1}]$ with a $U$-module structure via
\eq
&e_i(v^{(l)}_j\ot z^n)=e_iv^{(l)}_j\ot z^{n+{\delta_{i\,0}}},
\q f_i(v^{(l)}_j\ot z^n)=f_iv^{(l)}_j\ot z^{n-\delta_{i\,0}},\cr
&wt(v^{(l)}_j\ot z^n)=n\delta+(l-2j)\bar{\La}_1.&(ACT) \cr
\endeq
Namely $V_z^{(l)}$ is the affinization of $V^{(l)}$ \refto{(KMN)^2}.
Let $\{v_j^{(l)*}\}$ be a basis of $V^{(l)*}$ dual to $\{v^{(l)}_j\}$.
Then the following are isomorphisms of $U$-modules:
\eq
&C_{\pm}^{(l)}~:~V_{zq^{\mp2}}^{(l)}~
\buildrel{\sim}\over\longrightarrow ~V_z^{(l)*a^{\pm1}},\qquad
C_{\pm}^{(l)} v_j^{(l)}= (-1)^j q^{-j(l-j\mp 1)}
{\qbinom{l}{j}}^{-1} v^{(l)*}_{l-j}.
&(isotodual) \cr
\endeq
For $l=1$ they simplify to $C_\pm v_+=v_-^*$,
$C_\pm v_-=-q^{\pm 1} v_+^*$ where $C_\pm=C_\pm^{(1)}$.

Consider now an $R$ matrix
$R(z_1,z_2)\in {\hbox{End}}(V^{(m)}_{z_1}\ot V_{z_2}^{(n)})$
which intertwines the action of $U$ in the sense
\eq
&R(z_1,z_2)\Delta(x)=\Delta'(x)R(z_1,z_2)
\quad \forall x\in U. \cr
\endeq
These equations admit a unique solution up to scalar multiple.
We denote by $\ovl R_{mn}(z)$ the solution normalized as
\eq
&\ovl R_{mn}(z)\bigl(v^{(m)}_0\ot v^{(n)}_0\bigr)
=v^{(m)}_0\ot v^{(n)}_0.
\endeq
The following properties are relevant to the subsequent discussions.

\vskip 3mm

\noindent {\sl The inversion relations. \q}
\eq
&\ovl R_{mn}(z)P\ovl R_{nm}(z^{-1})P=1 \qbox{ where }
\q P(u\ot v)=v\ot u, &(Inv1)\cr
&\bigl(\ovl R_{mn}(z)^{-1}\bigr)^{t_1}
=\beta_{mn}(z) \bigl(C_+^{(m)}\ot\id\bigr)
\ovl R_{mn}(zq^{-2})
\bigl(C_+^{(m)}\ot\id\bigr)^{-1}, &(Inv2)\cr
&\beta_{mn}(z)=q^{-mn}
\prod_{j=0}^{n-1}{ 1-q^{m-n+2j}z \over 1-q^{-m-n+2j}z }. \cr
\endeq
The symbol ${}^{t_1}$ means the transposition in the first component,
i.e. we set
$(f\ot g)^{t_1}=f^t\ot g\in\End(V^*\ot W)$
for $f\in \End(V)$ and $g\in\End(W)$.
\vskip 3mm

\noindent {\sl The case $m=1$ or $n=1$.\q} If we write
\eq
&\ovl R_{1n}(z) \bigl(v_{\epsilon}\ot v^{(n)}_j\bigr)
 =\sum_{\epsilon'=\pm}
  v_{\epsilon'}\ot r^{1n}_{\epsilon'\epsilon}(z)v^{(n)}_j,\quad
\ovl R_{m1}(z) \bigl(v^{(m)}_j\ot v_{\epsilon}\bigr)
 =\sum_{\epsilon'=\pm}
  r^{m1}_{\epsilon' \epsilon}(z)v^{(m)}_j\ot v_{\epsilon'},\cr
\endeq
then we have
\eq
\pmatrix{
r^{1n}_{++}(z) & r^{1n}_{+-}(z) \cr
r^{1n}_{-+}(z) & r^{1n}_{--}(z) \cr
}
&=
{1\over q^{1+n/2}-z^{-1}q^{-n/2}}
\pmatrix{
qt_1^{1/2}-z^{-1}t_1^{-1/2} & (q-q^{-1})z^{-1}f_1t_1^{1/2} \cr
(q-q^{-1})e_1t_1^{-{1/2}}   & qt_1^{-1/2}-z^{-1}t_1^{1/2} \cr
},\cr
\pmatrix{
r^{m1}_{++}(z) & r^{m1}_{+-}(z) \cr
r^{m1}_{-+}(z) & r^{m1}_{--}(z) \cr
}
&=
{1\over zq^{m/2}-q^{-1-m/2}}
\pmatrix{
zt_1^{1/2}-q^{-1}t_1^{-1/2} & (q-q^{-1})zt_1^{1/2}f_1 \cr
(q-q^{-1})t_1^{-1/2}e_1     & zt_1^{-1/2}-q^{-1}t_1^{1/2} \cr
}.
\endeq
Here $t_1^{1/2}$ is understood to act on $v^{(l)}_j$ as
$q^{l/2-j}v^{(l)}_j$.
In fact only integral powers of $q$ appear in these formulas.

\vskip 3mm

Let us determine the image of the universal $R$ matrix \refeq{univR}
in $\End\bigl(V_{z_1}^{(m)}\ot V_{z_2}^{(n)}\bigr)$
\eq
&R^+_{mn}(z)=\pi_{V^{(m)}\ot V^{(n)}}\bigl({\cal R}'(z)\bigr). \cr
\endeq
{}From the uniqueness of the $R$ matrix we have
\eq
&R^+_{mn}(z)=\rho_{mn}(z)\ovl R_{mn}(z) &(image)
\endeq
with some scalar $\rho_{mn}(z)\in F[[z]]$.
{}From \refeq{Univ3} it follows that
$R^+_{mn}(z)$ satisfies the second inversion relation
\refeq{Inv2} with $\beta_{mn}(z)=1$. This leads to the
relation $\rho(z)\rho(zq^{-2})=\beta_{mn}(z)$, giving
\eq
&\rho_{mn}(z)
=q^{-mn/2}{(q^{m-n+2}z;q^4)_{\infty}(q^{-m+n+2}z;q^4)_{\infty}
\over (q^{m+n+2}z;q^4)_{\infty}(q^{-m-n+2}z;q^4)_{\infty} }.
&(scal)
\endeq
Here
$$
(z;p)_{\infty}=\prod_{j=0}^{\infty}(1-zp^j).
$$
The argument leading to \refeq{scal} is due to \refto{FR}.

\subsec(3.3| Vertex operators)
Let $\la$, $\mu$ be dominant integral weights of level $k$.
We set
$\Del_{\la}=(\la,\la+2\rho)/(2k+4)$.
By vertex operators (VOs) we mean the following types of formal series
\eq
\VVO(\mu V^{(l)},\la)(z)&=z^{\Del_{\mu}-\Del_{\la}}\VO(\mu V^{(l)},\la)(z),
\quad
\VO(\mu V^{(l)},\la)(z)
=\sum (\VO(\mu V^{(l)},\la))_{j,n}\ot v^{(l)}_jz^{-n},
&(VO1)\cr
\VVO(V^{(l)}\mu,\la)(z)&=z^{\Del_{\mu}-\Del_{\la}}\VO(V^{(l)}\mu,\la)(z),
\quad
\VO(V^{(l)}\mu,\la)(z)
=\sum v^{(l)}_jz^{-n}\ot(\VO(V^{(l)}\mu,\la))_{j,n},
&(VO2)\cr
\endeq
whose coefficients are the maps
\eq
&(\VO(\mu V^{(l)},\la))_{j,n},~ (\VO(V^{(l)}\mu ,\la))_{j,n}
{}~:~V(\mu)_{\xi} ~\goto{}~ V(\mu)_{\xi-(l-2j)\bar{\La_1}+n\del},
&(wtcomp)
\endeq
such that they are intertwiners of $U$-modules
\eq
&\VO(\mu V^{(l)},\la)(z)~:~V(\la)~\goto{}~V(\mu)\widehat{\ot} V^{(l)}_z,
&(intI)\cr
&\VO(V^{(l)} \mu,\la)(z)~:~V(\la)~\goto{}~V^{(l)}_z\widehat{\ot }V(\mu).
&(intII)\cr
\endeq
This means that, as a formal series, e.g. \refeq{intI} satisfy
\eq
&\VO(\mu V^{(l)},\la)(z)\circ x =\Delta(x)\circ \VO(\mu V^{(l)},\la)(z),
\qq \forall x\in U\cr
\endeq
where the action on $V^{(l)}_z$ is according to \refeq{ACT}.
In \refeq{intI,intII} we set
$M\widehat{\ot}N=\oplus_\xi\prod_\nu M_\nu\ot N_{\xi-\nu}$.
Henceforth we will not bother writing $\widehat{\phantom{\ot}}$.
Note that \refeq{wtcomp} implies
\eq
&(d\ot \id)\Phit_\la^{\mu V^{(l)}}(z)-\Phit_\la^{\mu V^{(l)}}(z)d
=-z{d\over dz}\Phit_\la^{\mu V^{(l)}}(z)
&(degree)
\endeq
and likewise for $\Phit_\la^{V^{(l)}\mu}(z)$.
By abuse of notation
we shall also write $\Phi_\la^{\mu V^{(l)}}(1)$ to mean the
intertwiner of $U'$-modules
$\sum (\VO(\mu V^{(l)},\la))_{j,n}\ot v^{(l)}_j:
V(\la)\goto{}\Vh(\mu)\ot V^{(l)}$ with $\Vh(\mu)=\prod_\nu V(\mu)_\nu$.
In \refto{DFJMN} we called \refeq{intI} a VO of type I,
\refeq{intII} of type II.

For a VO \refeq{VO1} we define its `leading term' $v_{lt}$ by
$$
\eqalignno{
&\VO(\mu V^{(l)},\la)(z)\ket{\la}=\ket{\mu}\ot v_{lt}+\cdots,&(LT1)\cr
}
$$
where $\cdots$  means terms of the form
$\ket{u}\ot v$ with $\ket{u}\in \bigoplus_{\xi\ne\mu}V(\mu)_{\xi}$.
Set
$$
V^{(l)\mu}_{\la}=\{v\in V^{(l)}\,|\,\la\,\equiv \,\mu+\wt v
{}~{\hbox{\rm mod }}\del,\,\,
e_i^{\langle h_i,\mu\rangle+1}v=0\q i=0,1\}.
$$
We have the following criterion about the existence of VOs.
\proclaim Proposition \prop{3.1} \refto{DJO}.
Associating $\VO(\mu V^{(l)},\la)(z)$ with
its leading term we have an isomorphism of vector spaces
$$
\{\hbox{ VOs }V(\la)\goto{}V(\mu)\ot V^{(l)}_z\}
{}~~\goto{\sim}~~ V^{(l)\mu}_{\la}.
$$
\par\noindent
Similar statement holds for the VOs \refeq{VO2}.

Let $\la=m\La_0+n\La_1$ ($m+n=k$), and
take $l=k$ for \refeq{VO1},  $l=1$ for \refeq{VO2}.
Then up to scalar the nontrivial VOs are the following ones:
\eq
&\VO(\mu \Vk,\la)\ket{\la}=\ket{\mu}\ot v_m^{(k)}+\cdots,
\qbox{if  } \mu=\sigma(\la)=n\La_0+m\La_1,
&(NORM.a)\cr
&\VO(\V1\mu ,\la)\ket{\la}=v_\mp \ot \ket{\mu}+\cdots,
\qbox{if  }\mu=\la_{\pm}=(m\mp 1)\La_0+(n\pm 1)\La_1 \in P_+.
&(NORM.b)\cr
\endeq
Note that in \refeq{NORM.a}
both $\mu$ and $v^{(k)}_m$ are uniquely determined from $\la$;
this is a general feature of perfect modules \refto{(KMN)^2}.
Henceforth we shall fix the normalization \refeq{NORM.a}, \refeq{NORM.b}
for the VOs.

The isomorphism \refeq{isotodual} gives rise to the VOs
\eq
&\Phit_\la^{\s(\la) V^{(k)*a^{\pm 1}}}(z)
=\al^{(k)}_{\pm}(\la) \bigl(\id\ot C^{(k)}_\pm\bigr)
\Phit_\la^{\mu V^{(1)}}(zq^{\mp 2}), &(dualVOI)\cr
&\Phit_\la^{V^{(1)*a^{\pm 1}} \mu }(z)
=\al^{(1)}_{\pm}(\la,\mu) \bigl(C^{(1)}_\pm\ot\id\bigr)
\Phit_\la^{V^{(1)} \mu }(zq^{\mp 2}), &(dualVOII)\cr
\endeq
where the normalization constants $\al^{(k)}_\pm(\la)$,
$\al^{(1)}_{\pm}(\la,\mu)$ are so chosen that
\eq
&\Phit_\la^{\mu V^{(k)*a^{\pm 1}}}(z)\ket{\la}
=\ket{\mu}\ot v^{(k)*}_n +\cdots
\qbox{if  } \mu=\sigma(\la)=n\La_0+m\La_1, \cr
&\Phit_\la^{V^{(1)*a^{\pm 1}} \mu }(z)\ket{\la}
=v^*_\epsilon\ot \ket{\mu}+\cdots
\qbox{if  }\mu=(m\mp \epsilon)\La_0+(n\pm \epsilon)\La_1 \in P_+.
\endeq
For $k=1$ we have
$$
\al^{(1)}_{\pm}(\la,\mu)=\cases{-q^{\mp1}&if $\mu=(m-1)\La_0+(n+1)\La_1$,\cr
                                 1      &if $\mu=(m+1)\La_0+(n-1)\La_1$.\cr}
$$
The fractional powers are included by the same rule, e.g.
$\Phi_\la^{V^{(1)*a^{\pm 1}} \mu }(z)
=z^{\Delta_\mu-\Delta_\la}$ \break
$\times\Phit_\la^{V^{(1)*a^{\pm 1}} \mu }(z)$.

\subsec(3.4| Explicit formulas for two point functions)
For our subsequent discussions, we need the following types of correlation
functions:
\eq
&\fun(\nu,\nu V_2^{(k)},\mu,z_2,\mu V_1^{(k)},z_1), &(!!!!)\cr
&\fun(\nu,V_2^{(1)}\nu,\mu,z_2,\mu \Vk_1,z_1), &(!!!)\cr
&\fun(\nu,\nu V_2^{(k)},\mu,z_2,V_1^{(1)}\mu,z_1), &(!!)\cr
&\fun(\nu,V_2^{(1)}\nu,\mu,z_2,V_1^{(1)}\mu,z_1).  &(!)\cr
\endeq
See Appendix 1, \refeq{I}--\refeq{IV} as for the convention.
Let
\eq
&z_1^{\Del_{\mu}-\Del_{\la}}z_2^{\Del_{\nu}-\Del_{\mu}}\Psi(z_1/z_2)
\endeq
be one of \refeq{!!!!}-\refeq{!}.
Thus, for instance,
$\Psi(z_1/z_2)=
\br{\Phit_\mu^{\nu \Vk_2}(z_2)\Phit_\la^{\mu \Vk_1}(z_1)}$
for \refeq{!!!!} (we omit writing $\bra{\nu}$, $\ket{\la}$).
Then $\Psi(z)$ is a formal power series in $z$ taking values in
$V_1^{(l_1)}\ot V_2^{(l_2)}$ with appropriate $l_1,l_2$,
and satisfies the $q$-KZ equation \refeq{2qKZ} where
$R^+(z)=R^+_{mn}(z)$ is given in \refeq{image,scal}.
Thanks to the perfectness of $\Vk$,
in the cases \refeq{!!!!}--\refeq{!!}
$\Psi(z)$ can be determined up to scalar by using
Proposition \refprop{reduc}.
Hence the $q$-KZ equation reduces to a scalar equation.
In the case \refeq{!} the $q$-KZ equation can be solved by
using basic hypergeometric functions (see Appendix 2).

We list the formulas for $\Psi(z)$ below.
Fix
\eq
&\la=m\La_0+n\La_1, \quad m,n\in\Z_{\ge0},~m+n=k \cr
\endeq
and set
\eq
\la_{\pm}&=\la\pm(\La_1-\La_0),\q \la_{\pm}^{\pm}=\la\pm2(\La_1-\La_0),
\q s={1\over 2(k+2)}, \q p=q^{2k+4}, \cr
\xi(z;a,b)&=
{(az;p,q^4)_{\infty}(a^{-1}bz;p,q^4)_{\infty} \over
(q^2az;p,q^4)_{\infty}(q^{-2}a^{-1}bz;p,q^4)_{\infty} }, \qq
(z;p,q)_{\infty}=\prod_{i,j\geq0}(1-zp^iq^j). \cr
&&(convention)
\endeq

\vskip 3mm

\noindent {\sl The case \refeq{!!!!}. \q}
In this case $\mu=\sigma(\la)$ and $\nu=\la$. We have
\eq
&\br{\Phit_{\s(\la)}^{\la \Vk_2}(z_2)\Phit_\la^{\s(\la) \Vk_1}(z_1)}
=\psi(z_1/z_2) \sum_{j=0}^k a_j(z_1/z_2) v_{k-j}^{(k)}\ot v_j^{(k)},
&(sol!!!!)\cr
&\psi(z)={(q^{2(k+2)}z;q^4)_\infty \over (q^4z;q^4)_\infty}, \cr
&a_j(z)=(-1)^{n+j} q^{j(k+n+1-j)}
\sum_{\max(0,j-m)\le i \le \min(n,j)}
{\qbinom{n}{i}}{\qbinom{m}{j-i}}q^{n-(k+2)i}z^{n-i}.\cr
\endeq

\vskip 3mm

\noindent {\sl The case \refeq{!!!}. \q}
$\Psi(z)$ is non-trivial when $\mu$ and $\nu$ are the following.
\eq
&\mu=\sigma(\la)=n\La_0+m\La_1,\q
\nu=\sigma(\la_\pm)=\wgt(n\pm 1,m\mp 1).\cr
\endeq
The solutions are
\eq
&\br{\Phit_{\s(\la)}^{\V1_2 \s(\la_\pm)}(z_2)\Phit_\la^{\s(\la) \Vk_1}(z_1)}\cr
&\qq\qq=
{(q^{5}z_1/z_2;q^4)_\infty \over (q^{3}z_1/z_2;q^4)_\infty }
\times\cases{
 (v^{(k)}_m \ot v_+ - q v^{(k)}_{m-1}\ot v_-)
	&\qbox{for $+$,}\cr
 (v^{(k)}_m \ot v_- -  (qz_1/z_2)v^{(k)}_{m+1}\ot v_+)
	&\qbox{for $-$.}\cr
}\qq &(sol!!!)
\endeq

\vskip 3mm

\noindent {\sl The case \refeq{!!}. \q}
$\Psi(z)$ is non-trivial when $\mu$ and $\nu$ are the following.
\eq
&\mu=\la_\pm=\wgt(m\mp 1,n\pm 1),\q
\nu=\sigma(\la_\pm)=\wgt(n\pm 1,m\mp 1),\qq&(00:\pm)\cr
\endeq
The solutions are
\eq
&\br{\Phit_{\la_\pm}^{\s(\la_\pm) \Vk_2}(z_2)\Phit_\la^{\V1_1 \la_\pm}(z_1)}\cr
&\qq\qq=
{{(qz_1/z_2;q^4)_{\infty}}\over{(q^{-1}z_1/z_2;q^4)_{\infty}}}
\times\cases{
(v_-\ot v^{(k)}_{m-1}-q^{-1}  v_+\ot v^{(k)}_m)
	&\qbox{for $+$,}\cr
(v_+\ot v^{(k)}_{m+1}-q^{-1}(z_1/z_2) v_-\ot v^{(k)}_m)
	&\qbox{for $-$.}\cr
}\qq
&(sol!!)
\endeq

\vskip 3mm

\noindent {\sl The case \refeq{!}. \q}
$\Psi(z)$ is non-trivial when $\mu$ and $\nu$ are the following.
\eq
&\mu=\la_\pm=(m\mp 1)\La_0+(n\pm 1)\La_1,\q\nu=\la,
&(?:\pm)\cr
&\mu=\la_\pm=(m\mp 1)\La_0+(n\pm 1)\La_1,
\q\nu=\la_\pm^\pm=(m\mp 2)\La_0+(n\pm 2)\La_1.
&(???:\pm)\cr
\endeq
The solutions are as follows.
\eq
&\br{\Phit_\mu^{\V1_2 \nu}(z_2)\Phit_\la^{\V1_1 \mu}(z_1)}/\xi(z_1/z_2;1,q^4)
\cr
&=\hhg((2n+4)s,2s;(2n+2)s;q^{-{2}}z_1/z_2)v_-\ot v_+\cr
&\qq-{{1-q^2}\over{q(1-q^{2n+2})}}
\hhg({1+2s},(2n+4)s;1+(2n+2)s;q^{-2}z_1/z_2)v_+\ot v_-
\qbox{for \refeq{?} with $+$}, \cr
&=\hhg({1-2ns},2s;1-(2n+2)s;q^{-2}z_1/z_2)v_+\ot v_-\cr
&\qq-{{1-q^2}
\over{q(1-pq^{-2n-2})}}
{{z_1}\over{z_2}}
\times\hhg({1+2s},{1-2ns};{2-(2n+2)s};q^{-2}z_1/z_2)v_-\ot v_+
\qbox{for \refeq{?} with $-$},\cr
&=v_\mp \ot v_\mp \qbox{for \refeq{???:\pm}},\cr
&&(sol!) \cr
\endeq
where $\hhg(a,b;c;z)=\hg(p^a,p^b;p^c;z)$ (see Appendix 2.).

\subsec(3.5| Commutation relations)
In this subsection we shall give the commutation relations of the
various vertex operators.

\vskip 3mm

\noindent {\sl The case \refeq{!!!!}. \q}
We modify the scalar factor of the $R$ matrix and define
\eq
&R_{kk}(z)=r_{kk}(z) \ovl R_{kk}(z), \q
r_{kk}(z)=z^{-k/2}
{(q^2z;q^4)_\infty (q^{2k+2}z^{-1};q^4)_\infty
\over
(q^2z^{-1};q^4)_\infty (q^{2k+2}z;q^4)_\infty
}. &(newRk)
\endeq
Then we have
\eq
&R_{kk}(z_1/z_2)
\Phi^{\la \Vk_1}_{\sigma(\la)}(z_1)
\Phi^{\sigma(\la) \Vk_2}_{\la}(z_2)
=
\Phi^{\la \Vk_2}_{\sigma(\la)}(z_2)
\Phi^{\sigma(\la) \Vk_1 }_{\la}(z_1).
&(COM!!!!)
\endeq
In fact, this equality can be verified for the matrix element
$\bra{\la}\cdots\ket{\la}$ by rewriting the $q$-KZ equation.
The equality for arbitrary matrix elements follows automatically
since both sides are intertwiners (the argument is the same
as in the proof of Proposition 6.1 of \refto{DFJMN}).

\vskip 3mm

\noindent {\sl The case \refeq{!!!}, \refeq{!!}. \q}
Using the formulas \refeq{sol!!!}, \refeq{sol!!}
and arguing similarly as above,
we obtain the following commutation relations.
\eq
&\comp(\V1_1\sigma(\la_\pm),\sigma(\la),z_1,\sigma(\la) \Vk_2,z_2)\cr
&\qq\qq=
\tau(z_1/z_2)^{-1}
\comp(\sigma(\la_{\pm}) \Vk_2,\la_{\pm},z_2,\V1_1\la_{\pm},z_1),
&(COM10)\cr
&\comp(\Vs1_1\sigma(\la_{\pm})
,\sigma(\la),z_1,\sigma(\la) \Vk_2,z_2)\cr
&\qq\qq=
\tau(z_1/z_2)~
\comp(\sigma(\la_{\pm})
\Vk_2,\la_{\pm},z_2,\Vs1_1\la_{\pm},z_1).&(COM11)\cr
\endeq
Here
\eq
&\tau(z)=z^{-1/2} {\Theta_{q^4}(qz) \over \Theta_{q^4}(qz^{-1})}
&(tau)
\endeq
with  $\Theta_p(z)$ being defined in Appendix 2, \refeq{Eq8}.

\vskip 3mm

\noindent {\sl The case \refeq{!}. \q}
This case is more complicated compared to the previous ones.
To derive the commutation relations we invoke the
connection formula \refeq{Eq4} for the basic hypergeometric functions.

Let
$R_{VV}(z)=-R_{11}(z)$ be defined in \refeq{newRk} with $k=1$ and set
\eq
R_{VV^*}(z)&=({\hbox{\rm id}}\ot C_-) R_{VV}(zq^{-2})
({\hbox{\rm id}}\ot C_-)^{-1}, \cr
R_{V^*V^*}(z)&=(C_-\ot C_-)R_{VV}(z)(C_-\ot C_-)^{-1}.
\endeq
(Our $R_{VV}(z)$ and $r_{11}(z)$ are
$R_{VV}(z)$ and  $-r_0(z)$ in \refto{DFJMN} respectively).
The inversion relations take the form
\eq
&R_{VV}(z)PR_{VV}(z^{-1})P=1, \cr
&\bigl(R_{VV}(z)^{-1}\bigr)^{t_1}=
-R_{VV^*}(z).
\endeq
Retaining the notations of \refeq{sol!}
we have the following commutation relations.
\eq
&\comp(\V1_1\nu,\mu,z_1,\V1_2\mu,z_2)
=R_{VV}(z_1/z_2)\sum_{\mu'}\comp(\V1_2\nu,\mu',z_2,\V1_1\mu',z_1)
\Boltz(\la,\mu,{\mu'},\nu,{{z_1}\over{z_2}}).\qq&(COM1)\cr
\endeq
Here
\eq
&\Boltz(\la,\mu,\mu',\nu,z)
=-z^{\Del_{\la}+\Del_{\nu}-\Del_{\mu}-\Del_{\mu'}-1/2}
{\xi(z^{-1};1,pq^4) \over \xi(z;1,pq^4)}\times
\BBoltz(\la,\mu,\mu',\nu,z),\cr
&\BBoltz(\la,\la_+,\la_+,\la,z)=
{\Theta_p(pq^2) \over \Theta_p(pq^{-2n-2})}
{\Theta_p(pq^{-2n-2}z) \over \Theta_p(pq^2z)}, \cr
&\BBoltz(\la,\la_+,\la_-,\la,z)=
q^{-1}\times
{\Gamma_p\bigl((2n+2)s\bigr) \Gamma_p\bigl((2n+2)s\bigr)
\over
\Gamma_p\bigl((2n+4)s\bigr) \Gamma_p\bigl(2ns\bigr)}
{\Theta_p(pz) \over \Theta_p(pq^2z)}, \cr
&\BBoltz(\la,\la_-,\la_+,\la,z)=
q^{-1}\times
{\Gamma_p\bigl(1-(2n+2)s\bigr) \Gamma_p\bigl(1-(2n+2)s\bigr)
\over
\Gamma_p\bigl(1-(2n+4)s\bigr) \Gamma_p\bigl(1-2ns\bigr)}
{\Theta_p(pz) \over \Theta_p(pq^2z)}, \cr
&\BBoltz(\la,\la_-,\la_-,\la,z)=
z^{-1}\times
{\Theta_p(pq^2) \over \Theta_p(q^{2n+2})}
{\Theta_p(q^{2n+2}z) \over \Theta_p(pq^2z)}, \cr
&\BBoltz(\la,\la_{\pm},\la_{\pm},\la_{\pm}^{\pm},z)=1,\cr
&\BBoltz(\la,\mu,\mu',\nu,z)=0\qbox{otherwise},\cr
\endeq
and $\Gamma_p(z)$ denotes the $q$-gamma function \refeq{Eq8}.

The same relations \refeq{COM1}
hold if we replace $\Phi_\la^{V\mu}$ by
$\Phi_\la^{V^{*a^{-1}}\mu}$ and $R_{VV}(z)$ by $R_{V^*V^*}(z)$.

We have further
\eq
&\comp(\V1_1\nu,\mu,z_1,V^{(1)*a^{-1}}_2\mu,z_2)
=R_{VV^*}({{z_1}/{z_2}})\sum_{\mu'}
\comp(V^{(1)*a^{-1}}_2\nu,\mu',z_2,\V1_1\mu',z_1)\cr
&
\qq\qq\qq\qq\qq\times\,(-q)^{\pm\del_{\la,\nu}\del_{\mu,\mu'}}
\Boltz(\la,\mu,\mu',\nu,{{z_1}\over{q^2z_2}}).&(COM5)\cr
\endeq

In the neighborhood of $|z_1/z_2|=1$ both side of
\refeq{COM1} are holomorphic, while those of \refeq{COM5} with $\la=\nu$
have a simple pole at $z_1=z_2$. The residues of
them are given by the following  formulas.
\eq
&{\hbox{\rm Res}}_{z_1=z_2}
\comp(\V1_1\la,\la_\pm,z_1,V^{(1)*a^{-1}}_2\la_\pm,z_2)
d\left({z_1\over z_2}\right)
=(v_+\ot v^*_++v_-\ot v^*_-)\ot g^\pm_{\la}\,{\hbox{\rm id}}_{V(\la)},\qq
&(RES1)\cr
\endeq
where
\eq
&g^{+}_{\la}=q^{-1}\xi(q^2;1,q^4)
{{(q^2;p)_{\infty}(q^{2n+4};p)_{\infty}}
\over{(p;p)_{\infty}(q^{2n+2};p)_{\infty}}},&(G+)\cr
&g^{-}_{\la}=\xi(q^2;1,q^4)
{{(q^2;p)_{\infty}(pq^{-2n};p)_{\infty}}
\over{(p;p)_{\infty}(pq^{-2n-2};p)_{\infty}}},&(G-)\cr
\endeq
and $\xi(z;a,b)$ is defined in \refeq{convention}.
\beginsection \S4. Vertex models

\subsec(4.1|Space of states)
We now turn to the discussion of the vertex model
associated with the spin $k/2$ representation $\Vk$ of $U'$.

Let ${\cal H}_k=\oplus_{\la\in P^0_k}V(\la)$.
We set
\eq
&\F=\End({\cal H}_k)=\oplus_{\la,\mu\in P^0_k}\F_{\la\mu}, &(total)\cr
&\F_{\la\mu}=\Hom\bigl(V(\mu),V(\la)\bigr)\simeq V(\la)\ot V(\mu)^{*a}.
&(Fixbc) \cr
\endeq
Strictly speaking we consider the completion of these spaces in the
topology of formal power series in $q$.
This is necessary in order to accomodate the action of VOs which
produce infinite sums.
This point is discussed in \sec(7) of \refto{DFJMN}.
We shall not go into the details here. The symbols $\otimes$, etc. are
to be understood appropriately.

We take \refeq{total} to be the mathematical definition of
the space of states of the vertex model.
As discussed in the Introduction it should replace the
usual na\"{\i}ve picture of the infinite tensor product space
$\cdots \ot \Vk\ot \Vk \ot \Vk \ot \cdots$,
taking into account the boundary conditions for the states.
We shall come to such an interpretation shortly.

The left action of $U$ on $\F_{\la\mu}$ is given by
\eq
xf&=\sum x_{(1)}\circ f\circ a\bigl(x_{(2)}\bigr) \cr
\endeq
where $\Delta(x)=\sum x_{(1)}\ot x_{(2)}$ for $x\in U$.
One can also endow the same underlying space with a structure of right
$U$-module by setting
\eq
fx&=\sum a^{-1}\bigl(x_{(2)}\bigr)\circ g\circ x_{(1)}. \cr
\endeq
We use the letter $\F^{\,r}_{\la\mu}$ to denote this right $U$-module.
In the case $\mu=\la$, $\F_{\la\la}$ has the unique canonical
element, i.e. the identity $\id_{V(\la)}$.
We call it the vacuum and denote it by $\vac_\la\in \F_{\la\la}$,
${}_\la\!\dvac\in \F^{\,r}_{\la\la}$.
There is a natural inner product \refto{DFJMN}
\eq
&\brak{f}{g}=
{\tr_{V(\la)}\bigl(q^{-2\rho}fg \bigr)
\over
\tr_{V(\la)}\bigl(q^{-2\rho} \bigr)}
\qquad f\in \F^{\,r}_{\la\mu},~g\in \F_{\mu\la}.&(inner)
\endeq
It enjoys the property
\eq
&\brak{fx}{g}=\brak{f}{xg} \qquad \forall x\in U.
\endeq

\subsec(4.2| Local structure)
Consider now the VO of type I, $\Phit_\la^{\s(\la) \Vk}(z)$ and
$\Phit_{\s(\la) \Vk}^\la(z)$.
We normalize the latter as
\eq
&\Phit_{\s(\la) \Vk}^\la(z)\bigl(\ket{\s(\la)}\ot v^{(k)}_m\bigr)
=\ket{\la}+\cdots, \qquad \la=m\La_0+n\La_1.
\endeq

\proclaim Proposition \prop{perfect}.
\eq
&\Phit_{\s(\la) \Vk}^\la(z)\circ\Phit_\la^{\s(\la) \Vk}(z)
=g_\la\times \id_{V(\la)},
&(inv)\cr
&\Phit_\la^{\s(\la) \Vk}(z)\circ \Phit_{\s(\la) \Vk}^\la(z)
=g_\la\times \id_{V\bigl(\s(\la)\bigr)\ot\Vk},
&(inv2)\cr
\endeq
where
\eq
&g_{m\La_0+n\La_1}=q^{mn} {\qbinom{k}{m}}
{(q^{2(k+1)};q^4)_\infty \over (q^2;q^4)_\infty},
\quad k=m+n. &(gfactor)
\endeq

\Proof
It is known \refto{DJO} that \refeq{inv} holds with some scalar $g_\la$.
This scalar can be calculated using the two point function
\refeq{sol!!!!} and \refeq{isotodual}.
Calculating similarly we find
\eq
&\bigl\langle
\Phit_{\la}^{\s(\la) \Vk_1}(z)\Phit_{\s(\la)}^{\la V^{(k)*a}_2}(z)
\bigr\rangle
=g_\la \sum_{j=0}^k v^{(k)}_j \ot v^{(k)*}_j. \cr
\endeq
Noting that the right hand side belongs to the trivial representation,
we easily see that
\eq
&\Phit_{\la}^{\s(\la) \Vk_1}(z)\Phit_{\s(\la)}^{\la V^{(k)*a}_2}(z)
=g_\la ~\id_{V(\la)}\ot
\left(\sum_{j=0}^k v^{(k)}_j \ot v^{(k)*}_j\right). \cr
\endeq
This is equivalent to \refeq{inv2}. \qed

The results \refeq{inv,inv2} tell that
$\Phit_{\la}^{\s(\la) \Vk}$ gives an isomorphism
(on the spaces being completed properly).
Iterating this we have for any $n$ an isomorphism
\eq
&\Phi^{(n)}_\la~
:~V(\la)~\goto{\sim}~ V\bigl(\s^n(\la)\bigr)\ot \bigl(\Vk\bigr)^{\ot n}.
&(openmap)
\endeq
This makes it possible to realize the space $\F_{\la\mu}$ in a way
which incorporates the `local structure'
$\Vk\ot \cdots \ot \Vk$ in between:
\eq
&\F_{\la\mu}
\simeq V(\la)\ot V(\mu)^{*a}~\goto{\sim}~
V\bigl(\s^n(\la)\bigr)\ot \bigl(\Vk\bigr)^{\ot n} \ot V(\mu)^{*a}.
&(open)
\endeq
One may do an analogous construction for the right half $V(\mu)^{*a}$.
The weight $\la$ (resp.$\mu$) in $\F_{\la\mu}$
signifies the boundary condition
to the left (resp. right) spacial infinity of the lattice.

\subsec(4.3|Transfer matrix)
Let us define the row transfer matrix of the vertex model
in our framework
\eq
&T(z)=T_{\la\mu}^{\s(\la)\s(\mu)}(z)
{}~:~\F_{\la\mu}~\goto{} ~\F_{\s(\la)\s(\mu)}.
&(rTM)
\endeq
By definition it is the composition of
\eq
&V(\la)\ot V(\mu)^{*a}
{}~\goto{\Phi(z)\ot\sid}~
V\bigl(\s(\la)\bigr) \ot \Vk_z \ot V(\mu)^{*a}
{}~\goto{\sid\ot \Phi^{*t}(z)}~
V\bigl(\s(\la)\bigr)\ot V(\s(\mu))^{*a}
\endeq
where $\Phi(z)=\Phit_\la^{\s(\la)V}(z)$ and
$\Phi^{*t}(z)$ is the transpose of
$\Phit_{\s(\mu)}^{\mu V^{*a^{-1}}}(z)$.
In terms of the weight components
\eq
&\Phit_\la^{\s(\la) V}(z)=\sum_j \Phi_j(z)\ot v^{(k)}_j, \cr
&\Phit_{\s(\mu)}^{\mu V^{*a^{-1}}}(z)
=\sum_j \Phi_j^*(z) \ot v^{(k)*}_j, \cr
\endeq
we have
\eq
&T(z)(f) =\sum_j \Phi_j(z)\circ f\circ \Phi_j^*(z)
\qbox{ for }~ f\in \Hom\bigl(V(\mu),V(\la)\bigr).&(TMcomp)
\endeq
We call $T=g_\la^{-1}T(1)$ the translation operator
where $g_\la$ is given in \refeq{gfactor} (see the proposition below).
{}From the property \refeq{degree}
we find that
$[d,T(z)]=-z{d\over dz}T(z)$, and hence
\eq
&T(z)=z^{-d}\circ T(1) \circ z^d.
\endeq
The following states that the vacuum is invariant under the transfer matrix
and the translation in particular.

\proclaim Proposition \prop{vacuum}.
\eq
&T_{\la\la}^{\s(\la)\s(\la)}(z)\vac_\la=
g_\la \vac_{\s(\la )}.
\endeq

\Proof
{}From the definition of the vacuum and the transfer matrix the assertion
is equivalent to
\eq
& \sum_j \Phi_j(z) \circ \Phi_j^*(z)=g_\la\times \id
\endeq
in the notation of \refeq{TMcomp}.
This can be shown similarly as in the proof of Proposition \refprop{perfect}.
\qed

Let us consider the action of $T(z)$ in the picture \refeq{open}.
We set
\eq
&T^{(n)}(z)=\bigl(\Phi_{\s(\la)}^{(n)}\ot \id\bigr)
\circ T(z)\circ
\bigl({\Phi_\la^{(n)}} \ot \id\bigr)^{-1}, \cr
&T^{(n)}(z)~:~
V\bigl(\s^n(\la)\bigr)\ot V^{(k)\ot n}\ot V(\mu)^{*a}
{}~\goto{}~
V\bigl(\s^{n+1}(\la)\bigr)\ot V^{(k)\ot n}\ot V(\mu)^{*a}.
&(TM')
\endeq
Denote by
$\Tau(z)$ the monodromy matrix
on a lattice of horizontal length $n$
\eq
&\Tau(z)=R^{0n}(z)\cdots R^{02}(z)R^{01}(z)
\in \End(\Vk_0\ot \Vk_1\ot \cdots \ot \Vk_n)
\endeq
where $R^{ij}(z)$ signifies the $R$ matrix
$R(z)=R_{kk}(z)\in \End(\Vk\ot \Vk)$ \refeq{newRk}
acting on the $i$-th and $j$-th components.
Let $\Tau(z)_j^{j'}$ be its matrix elements given by
\eq
&\Tau(z)\bigl(v^{(k)}_j\ot w\bigr)
=\sum_{j'} v^{(k)}_{j'}\ot \Tau(z)_j^{j'}w .
\endeq
Usual definitioin of the row transfer matrix
(in the periodic boundary condition) is $\sum_j\Tau(z)_j^j$.
The following motivates our definition of the row transfer matrix
\refeq{rTM}.

\proclaim Proposition \prop{rowTM}.
In the picture \refeq{open} the row transfer matrix \refeq{rTM} acts
as follows:
\eq
&T^{(n)}(z)\bigl(u\ot v\ot w)=
\sum \Phi_j(z)u\ot \Tau(z)_j^{j'}v \ot \Phi^{*t}_{j'}(z)w \cr
&u\ot v\ot w \in V(\s^n(\la))\ot V^{(k)\ot n} \ot V(\mu)^{*a}.
\endeq
Here $\Phi^{*t}_{j'}(z)$ signifies the transpose of $\Phi^{*}_{j'}(z)$.

\Proof
Using the commutation relation \refeq{COM!!!!} repeatedly we have
\eq
&\Tau(z)\, \Phit^{\s^{n+1}\la V_0}_{\s^n\la}(z)\, \Phi^{(n)}_\la
=\Phi_{\s(\la)}^{(n)}\, \Phit^{\s\la V_0}_{\la}(z),
\endeq
or in components
\eq
&\Phi_{\s(\la)}^{(n)} \Phi_{j'}(z) \bigl(\Phi_\la^{(n)}\bigr)^{-1}
=\sum_j \Phi_j(z)\Tau(z)_j^{j'}.
\endeq
The assertion follows from this,
\refeq{TMcomp} and \refeq{TM'}.
\qed

\def\cre(#1,#2,#3,#4){\varphi^{*{#1}}_{#3,{#2}}(#4)}
\def\ann(#1,#2,#3,#4){\varphi^{ {#1}}_{#3,{#2}}(#4)}
\def\gen(#1,#2,#3){\psi^{*{#1}}_{{#2},{#3}}}
\def\ep{\epsilon}
\def\inp(#1,#2){(#1;#2)_{\infty}}

\subsec(4.4|Creation and annihilation operators)
In the same way as \refto{DFJMN}, the particles are created over the
vacuum by the type II VOs.
Let us define their components by
\eq
&\Phi^{\V1\mu}_{\la}(z)=v_+\ot \Phi^{\mu}_{\la,+}(z)
+v_-\ot \Phi^{\mu}_{\la,-}(z), &(anncomp)\cr
&\Phi^{V^{(1)*a^{-1}}\mu}_{\la}(z)=v^*_+\ot \Phi^{*\mu}_{\la,+}(z)
+v_-^*\ot \Phi^{*\mu}_{\la,-}(z). &(crecomp)\cr
\endeq
{}From \refeq{dualVOII} we have the relation
\eq
&\Phi_{\la,+}^{\la_+}(z)=(-q)^{-1}\Phi_{\la,-}^{*\la_+}(zq^{-2}),
\q
\Phi_{\la,-}^{\la_+}(z)=\Phi_{\la,+}^{*\la_+}(zq^{-2}),\cr
&\Phi_{\la,+}^{\la_-}(z)=\Phi_{\la,-}^{*\la_-}(zq^{-2}),
\q
\Phi_{\la,-}^{\la_-}(z)=(-q)\Phi_{\la,+}^{*\la_-}(zq^{-2}).\cr
\endeq

Define the creation operator $\cre(\la',\ep,\la,z)$ on
$\F_{\la\mu}=\Hom\bigl(V(\mu),V(\la)\bigr)$
by
\eq
&\cre(\la',\ep,\la,z)~:~\F_{\la\mu}~\goto{}~\F_{\la'\mu},
\quad f~\mapsto ~\Phi^{*\la'}_{\la,\ep}(z) \circ f.
\endeq
Likewise define the annihilation operator
$\ann(\mu',\ep,\mu,z)$ to be the adjoint (with respect to the inner product
\refeq{inner}) of
\eq
&\ann(\mu',\ep,\mu,z)~:~\F_{\la\mu'}^r~\goto{}~\F_{\la\mu}^r,
\quad f~\mapsto ~f \circ \Phi^{\mu'}_{\mu,\ep}(z).
\endeq
In both cases the quasi-momentum $z$ is supposed to be on the unit
circle $|z|=1$.

{}From \refeq{COM1} and \refeq{COM5}
we obtain the following commutation relations of
creation and annihilation  operators.
\eq
&\ann(\nu,\ep_1,\mu,z_1)\ann(\mu,\ep_2,\la,z_2)
=\sum_{\mu',\ep_1',\ep_2'}
{R_{VV}({{z_1}/{z_2}})}_{\ep_1\ep_2}^{\ep'_1\ep'_2}~
\ann(\nu,\ep'_2,\mu',z_2)\ann(\mu',\ep'_1,\la,z_1)
\Boltz(\la,\mu,\mu',\nu,{{z_1}\over{z_2}}),\qq
&(AA1)\cr
&\cre(\nu,\ep_1,\mu,z_1)\cre(\mu,\ep_2,\la,z_2)
=\sum_{\mu',\ep_1',\ep_2'}
{R_{V^*V^*}({{z_1}/{z_2}})}_{\ep_1\ep_2}^{\ep'_1\ep'_2}~
\cre(\nu,\ep'_2,\mu',z_2)\cre(\mu',\ep'_1,\la,z_1)
\Boltz(\la,\mu,\mu',\nu,{{z_1}\over{z_2}}),\qq
&(CC1)\cr
&\ann(\nu,\ep_1,\mu,z_1)\cre(\mu,\ep_2,\la,z_2)
=\sum_{\mu',\ep_1',\ep_2'}
{R_{VV^*}({{z_1}/{z_2}})}_{\ep_1\ep_2}^{\ep'_1\ep'_2}~
\cre(\nu,\ep'_2,\mu',z_2)\ann(\mu',\ep'_1,\la,z_1)\cr
&\qq\qq\qq\qq\times\,(-q)^{\pm\del_{\la,\nu}\del_{\mu,\mu'}}
\Boltz(\la,\mu,\mu',\nu,{{z_1}\over{q^2z_2}})
+g_{\la}^{\pm}\del_{\la,\nu}\del_{\ep_1,\ep_2}\del(z_1/z_2).
&(AC1)\cr
\endeq
Here we have set
$R_{VV}(z)v_{\ep_1}\ot v_{\ep_2}=\sum
v_{\ep'_1}\ot v_{\ep'_2}R_{VV}(z)^{\ep'_1,\ep'_2}_{\ep_1,\ep_2} $.
The delta function $\del(z)=\sum_{n\in {\bf Z}}z^n$ in \refeq{AC1}
arises because of the pole of \refeq{COM5} at $z_1=z_2$.

To find the commutation relation with the transfer matrix $T(z)$ we use
\refeq{COM11} to get
\eq
&T_{\la'\mu}^{\s(\la')\s(\mu)}(z_2)\Phi^{*\la'}_{\la,\epsilon}(z_1)
=\tau(z_1/z_2)^{-1}
\Phi^{*\s(\la')}_{\s(\la),\epsilon}(z_1)T_{\la\mu}^{\s(\la)\s(\mu)}(z_2),
&(Tphi)
\endeq
where $\tau(z)$ is given in \refeq{tau}.

\def\Tt{\widetilde{T}}
Let
$\Tt(z)=T^{\la\mu}_{\s(\la)\s(\mu)}(z)T_{\la\mu}^{\s(\la)\s(\mu)}(z)$
be the transfer matrix for two lattice steps
which does not change the boundary conditions.
Then \refeq{Tphi} implies
\eq
&\Tt(z_2)\ann(\mu,\ep,\la,z_1)\Tt(z_2)^{-1}
=\tau(z_1/z_2)^2 \ann(\mu,\ep,\la,z_1), \cr
&\Tt(z_2)\cre(\mu,\ep,\la,z_1)\Tt(z_2)^{-1}
=\tau(z_1/z_2)^{-2} \cre(\mu,\ep,\la,z_1). \cr
\endeq
Note that the function $\tau(z)$ is independent of the level $k$
and therefore is the same as (7.11) in \refto{DFJMN}.
The energy can be obtained by differentiating $\log\tau(z)$.
Hence our formulas agree with Sogo's result \refto{Sogo}
found through the Bethe Ansatz method.

\def\Lc{{\cal L}}
\def\Tr{\hbox{Tr}}

\subsec(4.5|Local operators)
The local operators can be defined making use of the realzation
\refeq{open}.
For a linear map $L\in \End\bigl(V^{(k)\ot n}\bigr)$
we define its action
on $\F_{\la\mu}$ by $f\mapsto \Lc_\la\circ f$ where we set
\refto{JMMN}
\eq
&\Lc_\la = \Phi_\la^{(n)-1}\bigl(\id\ot L\bigr)\Phi_\la^{(n)}
\endeq
in the notation of \refeq{openmap}.
Using \refeq{COM10,COM11} we find that the components of the
creation operator \refeq{crecomp} commutes with $\Lc_\la$ in the sense
\eq
&\Lc_\mu \Phi^{*\mu}_{\la\ep}(z)
=\Phi^{*\mu}_{\la\ep}(z)\Lc_\la.
\endeq

For simplicity of notation,  we shall write in the sequel
\eq
&V^*=V^{(1)*a^{-1}}, \quad
R(z)=R_{kk}(z), \quad R^*(z)=R_{V^*V^*}(z), &(conv)\cr
&\Tr(\cdot)=\tr(\cdot)/\tr(q^{-2\rho}).
\endeq
Consider the $n$-particle states created from the vacuum
\eq
\varphi^{*\s^n(\la)}_{\s^{n-1}(\la),\ep_n}(z_n)\cdots
\varphi^{*\s(\la)}_{\la,\ep_1}(z_1)\vac_\la=
&\Phi^{*\s^n(\la)}_{\s^{n-1}(\la),\ep_n}(z_n)\cdots
\Phi^{*\s(\la)}_{\la,\ep_1}(z_1) \cr
&\qquad \in
\Hom\bigl(V(\la),V(\s^n(\la))\bigr)
&(n-state)
\endeq
where $\ep_1,\cdots,\ep_n\in \{\pm\}$.
A form factor of $L\in \End\bigl(V^{(k)\ot n}\bigr)$
is  defined by (we put $\la_n=\la_0$)
\eq
&F_{\la_n,\cdots,\la_0}(z_n,\cdots,z_1)
=\Tr_{V(\la_0)}\Bigl(q^{-2\rho}\Lc_{\la_0} \Phi_{\la_{n-1}}^{V_n^*\la_0}(z_n)
\cdots \Phi_{\la_0}^{V_1^*\la_1}(z_1)\Bigr).
&(Formfac)
\endeq
It is a  $V_1^*\ot \cdots \ot V_n^*$-valued function where
 $V^*_j$ is a copy of $V^*$, the suffix $j$ indicating
that it is put at the $j$-th place in the tensor product.
More generally let us consider the following trace
\def\zt{\zeta}
\eq
&F_{\la_n,\cdots,\la_0}(\zt_1,\cdots,\zt_m,z_n,\cdots,z_1) \cr
&=\Tr_{V(\la)}
\Bigl(q^{-2\rho}\Phi^{\la\Vk_1}_{\s(\la)}(\zt_1)\cdots
\Phi^{\s^{m-1}(\la)\Vk_m}_{\s^m(\la)}(\zt_m)
\Phi^{V^*_n \la_n}_{\la_{n-1}}(z_n)\cdots
\Phi^{V^*_1 \la_1}_{\la_0}(z_1)
\Bigr)&(trace)
\endeq
where $\la_0=\la$ and $\la_n=\s^m(\la)$.
It is a function with values in $V_1^*\ot \cdots \ot V_n^*\ot
\Vk_1\ot \cdots \ot \Vk_m$.

\proclaim Proposition \prop{Formfactor}.
The following difference equations hold for \refeq{trace}:
\eq
&F_{\la_n,\cdots,\la_0}(\zt_1,\cdots,q^4\zt_j,\cdots,\zt_m,z_n,\cdots,z_1)
/\prod_{i=1}^n\tau(q^4\zt_j/z_i)  \cr
&=~R^{jj+1}(q^4\zt_j/\zt_{j+1})^{-1} \cdots R^{j m}(q^4\zt_j/\zt_m)^{-1} \cr
&\quad\times \pi_{\Vk_j}\bigl(q^{-2\bar{\rho}}\bigr)
R^{1j}(\zt_1/\zt_j) \cdots R^{j-1 j}(\zt_{j-1}/\zt_j)
F_{\s(\la_n),\cdots,\s(\la_0)}(\zt_1,\cdots,\zt_m,z_n,\cdots,z_1), \cr
&
F_{\la_n,\cdots,\la_0}(\zt_1,\cdots,\zt_m,z_n,\cdots, q^4z_i,\cdots,z_1)
/\prod_{j=1}^m\tau(z_i/\zt_j)~ \cr
&=R^{* i-1i}(z_{i-1}/q^4 z_i)^{-1} \cdots R^{*1i}(z_1/q^4 z_i)^{-1} \cr
&\quad\times
\pi_{V^*_i}\bigl(q^{-2\bar{\rho}}\bigr)
R^{*in}(z_i/z_n) \cdots R^{*ii+1}(z_i/z_{i+1}) \cr
&\sum_{\la'_n,\cdots,\la'_{i+1},\la'_{i-1},\cdots,\la'_1}
F_{\s^m(\la'_1),\la'_n,\cdots,\la_i,\cdots,\la'_1}
(\zt_1,\cdots,\zt_m,z_n,\cdots,z_1) \cr
&\times
\Boltz(\la_0,\la_1,\la'_1,\la'_2,q^4z_i/z_1)
\cdots
\Boltz(\la_{i-2},\la_{i-1},\la'_{i-1},\la_i,q^4z_i/z_{i-1}) \cr
&\times
\Boltz(\la_i,\la_{i+1},\la'_{i+1},\la'_{i+2},z_i/z_{i+1})
\cdots
\Boltz(\la_{n-1},\la_n,\la'_n,\s^m(\la'_1),z_i/z_n).
\endeq
Here $\pi_{\Vk_j}\bigl(q^{-2\bar{\rho}}\bigr)$
(resp. $\pi_{V^*_i}\bigl(q^{-2\bar{\rho}}\bigr)$)
signifies the operator $q^{-2\bar{\rho}}$ acting on
the component $V_j$ (resp. $V^*_i$).
The notation $R^{ij}(z)$ means the $R(z)$ acting on $V_i\ot V_j$
and likewise for $R^{*ij}(z)$ (not to be confused with
the $R_{kk}(z)$ in \refeq{conv}).

\Proof
Note that
\eq
&\Phi_\la^{V^*\mu}(q^4z)=
\bigl(q^{-2{\bar{\rho}}}\ot q^{-2\rho}\bigr)
\Phi_\la^{V^*\mu}(z)q^{2\rho}, \cr
&\Phi_\la^{\s(\la)\Vk}(q^4z)=
\bigl(q^{-2\rho} \ot q^{-2\bar{\rho}} \bigr)
\Phi_\la^{\s(\la) \Vk}(z)q^{2\rho}.
\endeq
The assertions follow from this, the commutation relations
\refeq{COM!!!!,COM10,COM11,COM1}
and the cyclic property of the trace.
\qed

In particular, repalcing the $\zt_i$ by
$(q^{-2}\zt_1,\cdots,q^{-2}\zt_m, \zt_m,\cdots,\zt_1)$
and using \refeq{dualVOI} we obtain
the following relation for arbitrary form factors
\eq
&F_{\la_n,\cdots,\la_0}(z_n,\cdots,q^4z_i,\cdots,z_0)  \cr
&=
R^{*i-1i}(z_{i-1}/q^4 z_i)^{-1} \cdots R^{*1i}(z_1/q^4 z_i)^{-1} \cr
&\quad \times \pi_{V^*_i}\bigl(q^{-2\bar{\rho}}\bigr)
R^{*in}(z_i/z_n) \cdots R^{*ii+1}(z_i/z_{i+1}) \cr
&\sum_{\la'_n,\cdots,\la'_{i+1},\la'_{i-1},\cdots,\la'_1}
F_{\la'_1,\la'_n,\cdots,\la_i,\cdots,\la'_1}(z_n,\cdots,z_1) \cr
&\times
\Boltz(\la_0,\la_1,\la'_1,\la'_2,q^4z_i/z_1)
\cdots
\Boltz(\la_{i-2},\la_{i-1},\la'_{i-1},\la_i,q^4z_i/z_{i-1}) \cr
&\times
\Boltz(\la_i,\la_{i+1},\la'_{i+1},\la'_{i+2},z_i/z_{i+1})
\cdots
\Boltz(\la_{n-1},\la_n,\la'_n,\la'_1,z_i/z_n).
\endeq

In the case of level $k=1$ there is only one term in the sum.
This is then a lattice analog of the $q$KZ equation
found by Smirnov \refto{Smbk} for form factors in massive
integrable field theories.

\par
\bigskip\noindent
{\it Acknowledgement.}\quad
We wish to thank
E. Date,
H. de Vega,
K. Miki,
A. Nakayashiki,
M. Okado
and
E. K. Sklyanin
for discussions and interest in this work.
\numberby{\beginsection}\prefixby{A}

\def\VO(#1,#2){{\tilde \Phi}^{#1}_{#2}}
\def\h(#1){|u_{#1}\rangle}

\def\lv{\langle{\rm vac}|}

\def\fun(#1,#2,#3,#4,#5,#6){\bra{#1}\Phi^{#2}_{#3}(#4)
                            \Phi^{#5}_{\la}(#6)\ket{\la}}
\def\ffun(#1,#2,#3,#4,#5,#6,#7,#8){\bra{#1}\Phi^{#2}_{#3}(#4){#5}
                            \Phi^{#6}_{\la}(#7){#8}\ket{\la}}
\def\fffun(#1,#2,#3,#4,#5,#6,#7,#8){\bra{#1}{#2}\Phi^{#3}_{#4}(#5)
                            \Phi^{#6}_{\la}(#7){#8}\ket{\la}}

\def\q{\quad}
\def\qq{\qquad}
\def\pair(#1,#2){{\langle{#1},{#2}\rangle}}
\def\al{\alpha}
\def\ot{\otimes}
\def\V(#1,#2){V^{(#1)}(#2)}
\def\ovl{\overline}
\def\del{\delta}
\def\Del{\Delta}
\def\hg(#1,#2;#3;#4){{}_2\phi_1\left({{{#1}\,\,\,{#2}}\atop{{#3}}};q,
                      {#4}\right)}
\def\conn(#1,#2,#3){{\Gamma_q({#1})\Gamma_q({#2})\Theta_q({#3})}}

\beginsection Appendix 1. The $q$-KZ equation

In this appendix we briefly outline the derivation of the $q$-KZ equation
for the correlation functions of vertex operators \refto{FR}.
We shall not go into the discussions about the topology here.

\subsec(A1.1| Universal $R$ and Drinfeld's central element)
Let $U=\uqg$ be the quantized enveloping algebra associated with
an affine Lie algebra $\goth{g}$.
Concerning the notations of affine Lie algebras
we follow \refto{Kac}, except that we reverse the ordering of
vertices for the type $A^{(2)}_{2l}$.

Recall that $U$ has the universal $R$ matrix
$\cal R$ that enjoys the following properties \refto{D, Ta}.
\eq
&{\cal R}\Delta(x)=\Delta'(x){\cal R}\qbox{for any}\q x\in U, &(Univ1)\cr
&(\Delta\ot{\hbox{\rm id}}){\cal R}
 ={\cal R}_{13}{\cal R}_{23},\q
 ({\hbox{\rm id}}\ot\Delta){\cal R}
 ={\cal R}_{13}{\cal R}_{12},&(Univ2)\cr
&(\varepsilon\ot{\hbox{\rm id}}){\cal R}
 =1=({\hbox{\rm id}}\ot\varepsilon){\cal R},\q
 (a\ot{\hbox{\rm id}}){\cal R}
 ={\cal R}^{-1}=({\hbox{\rm id}}\ot a^{-1}){\cal R}. &(Univ3)
\endeq
Here $\varepsilon$ is the counit, $a$ is the antipode,
$\Delta'=\sigma\circ\Delta$ with $\sigma(a\ot b)=b\ot a$, and
if we write ${\cal R}=\sum a_i\ot b_i$ then
${\cal R}_{12}=\sum a_i\ot b_i\ot1$,
${\cal R}_{23}=\sum 1\ot a_i\ot b_i$,
${\cal R}_{13}=\sum a_i\ot 1\ot b_i$.

Let $\{\ovl h_i\}$, $\{\ovl h^i\}$ be the dual bases of the classical part
of the Cartan subalgebra of $\goth{g}$,
and let $c$, $d$ be
the canonical central element and the scaling element respectively.
Let further
$u_{\beta i}$, $u^i_{-\beta}$ be dual bases of the weight $\pm\beta$
component of the `nilpotent' subalgebras (see \refto{Ta}).
We set
\eq
&{\cal R}'(z)=q^{-\sum \ovl{h_i}\ot\ovl{h^i}}
\sum_{\beta,i} z^{\langle d,\beta\rangle}u_{\beta\, i}\ot u^i_{-\beta}.
\endeq
Note that if $\beta=\sum_{i=0}^nm_i\alpha_i$, then
$\langle d,\beta\rangle=m_0$.
The universal $R$ matrix iteslf is given by
\eq
{\cal R}&=q^{-c\ot d-d\ot c}{\cal R}', \quad
{\cal R}'={\cal R}'(1)
=q^{-\sum \ovl{h_i}\ot\ovl{h^i}}\sum u_{\beta i}\ot u^i_{-\beta}. \cr
\endeq
In terms of $\cal R'$ the property \refeq{Univ2} is translated as follows.
\eq
&(\Delta\ot{\hbox{\rm id}})({\cal R}')
={\cal R}'_{13}(q^{-c_2}){\cal R}'_{23}, \quad
({\hbox{\rm id}}\ot\Delta)({\cal R}')
={\cal R}'_{13}(q^{c_2}){\cal R}'_{12}, &(Univ')\cr
\endeq
where $c_2=1\ot c\ot 1$.
\par

The following are shown in \refto{D2}.
\item{(i)} If ${\cal R}=\sum a_i\ot b_i$, then
${\cal C}=\sum a(b_i)a_i$ is invertible and
${\cal C}^{-1}=\sum a^{-1}(b_i)a(a_i)$,
\item{(ii)} ${\cal C}x{\cal C}^{-1}=a^2(x)$ $\forall x\in U$,
\item{(iii)} The element ${\cal D}=q^{2\rho}{\cal C}$
($\rho=\sum \La_i$) is central in certain completion of $U$,
\item{(iv)} $\cal D$ acts on $V(\la)$ as a scalar
$q^{(\la,\la+2\rho)}$. Here $V(\la)$ denotes the irreducible $U$-module
with highest weight $\la$.

\subsec(A1.2| Difference equaitons for VO)
We shall study how the central element $\cal D$
behaves with respect to intertwiners.

\proclaim Proposition \prop{A}.
Let $V_1$, $V_2$ and $V_3$ be $U$-modules, and let
$\Phi:V_1\longrightarrow V_2\ot V_3$ be a $U$-linear map. Then we have
\eq
&({\cal D}\ot{\hbox{\rm id}})\Phi{\cal D}^{-1}
=\sigma({\cal R})\sum(1\ot b_iq^{-2\rho})\Phi a_i, &(*)\cr
&({\hbox{\rm id}}\ot {\cal D})\Phi{\cal D}^{-1}
=\sum(b_iq^{-2\rho}\ot 1){\cal R}\Phi a_i. &(***)  \cr
\endeq

\Proof
The derivation being similar, we show \refeq{*}.
We shall use the sigma notation $\Delta(x)=\sum x_{(1)}\ot x_{(2)}$,
$\bigl(\Delta\ot 1\bigr)\Delta(x)=\sum x_{(1)}\ot x_{(2)}\ot x_{(3)}$,
etc. for $x\in U$.

First note that the intertwining property of $\Phi$ is equivalently stated
as follows.
\eq
\sum \bigl(x_{(1)}\ot x_{(2)}\bigr)  \Phi=\Phi  x
&\Longleftrightarrow
\sum \bigl(a(x_{(1)})\ot 1\bigr)  \Phi  x_{(2)}
= \bigl(1\ot x\bigr)  \Phi  &(Hopf1)\cr
&\Longleftrightarrow
\sum \bigl(1\ot a^{-1}(x_{(2)})\bigr)  \Phi x_{(1)}
=\bigl( x\ot 1 \bigr)  \Phi. &(Hopf2)\cr
\endeq
Using these we have
\eq
\bigl({\cal C}\ot 1\bigr) \Phi
&=\sum \bigl(a(b_i)a_i \ot 1\bigr) \Phi \cr
&=\sum \bigl(a(b_i)\ot a^{-1}(a_{i(2)})\bigr) \Phi a_{i(1)} \cr
&=\sum \bigl(a(b_ib_j) \ot a^{-1}(a_j)\bigr) \Phi a_i. \cr
\endeq
In the last step we used the fact that
$\sum a_{i(1)}\ot a_{i(2)}\ot b_i =\sum a_i\ot a_j\ot b_ib_j$,
which follows from the first of \refeq{Univ2}.
Similarly
\eq
\Phi {\cal C}^{-1}
&=\Phi \sum a^{-1}(b_r)a(a_r) \cr
&=\sum \bigl(a^{-1}(b_{r(2)})\ot a^{-1}(b_{r(1)})\bigr)\Phi a(a_r) \cr
&=\sum \bigl(a^{-1}(b_r)\ot a^{-1}(b_s)\bigr) \Phi a(a_ra_s).
\endeq
Combining them we have
\eq
{\cal C}\ot 1 \Phi{\cal C}^{-1}
&=\sum \bigl(a(b_ib_j) \ot a^{-1}(a_j)\bigr) \Phi a_i {\cal C}^{-1} \cr
&=\sum \bigl(a(b_ib_j) \ot a^{-1}(a_j)\bigr)
\bigl(a^{-1}(b_r)\ot a^{-1}(b_s)\bigr) \Phi
a(a_ra_s){\cal C}a_i {\cal C}^{-1} \cr
&=\sum \bigl(a(b_j)a(b_i)a^{-1}(b_r) \ot a^{-1}(a_j)a^{-1}(b_s)\bigr)
\Phi
a(a_s)a(a_r)a^2(a_i). \cr
\endeq
On the  other hand, \refeq{Univ3} implies
$\sum_i a^2(a_i)\ot a(b_i)=\sum_i a(a_i)\ot b_i$ and
$\sum_{i\,r}a_ia_r\ot b_ia^{-1}(b_r)=1$, hence
the right hand side can be simplified to
\eq
&\sum \bigl(a(b_j)\ot a^{-1}(b_sa_j)\bigr) \Phi a(a_s)
=\sum \bigl(a^2(b_j)\ot a_jb_s\bigr) \Phi a^2(a_s).
\endeq
The assertion follows easily from this.
\qed

Let $V$ be a finite dimensional module over the subalgebra
$U'$ `with $d$ being dropped', and let $V_z$ denote its affinization.
We shall apply Proposition \refprop{A} to the $U$-linear maps of the form
$\tilde\Phi_1(z):V(\la)\longrightarrow V(\mu)\widehat{\ot}V_z$
and $\tilde\Phi_2(z):V(\la)\longrightarrow V_z\widehat{\ot}V(\mu)$.
(Here $M\widehat{\ot}N=\sum_\xi \prod_\nu M_\nu\ot N_{\xi-\nu}$.)
Set
\eq
&p=q^{2(k+h^{\vee})}, \qquad
\Delta_{\la}={{(\la,\la+2\rho)} \over {2(k+h^{\vee})}}
\endeq
where $k$ is the level of $\la$
and $h^{\vee}$ denotes the dual Coxeter number.
Writing ${\cal R}'=\sum a'_i\ot b'_i$ we rephrase
Proposition \refprop{A} as
\eq
&({\cal D}\ot {p^d})\tilde\Phi_1(z){\cal D}^{-1}
=\sum \Bigl(b'_i\ot {\hbox{\rm Ad}}\bigl(q^{(k+2h^{\vee})d}\bigr)a'_i\Bigr)
\Bigl(1\ot q^{-2\ovl{\rho}}a^{-2}(b'_j)\Bigr)\tilde\Phi_1(z)a'_j,&(DD1)\cr
&(p^d\ot {\cal D})\tilde\Phi_2(z){\cal D}^{-1}
=\sum \Bigl(q^{-2\ovl\rho}\bigl({\hbox{\rm Ad}}(q^{kd})a^{-2}(b'_j)\bigr)
a'_i\ot b'_i\Bigr)
\tilde\Phi_2(z)a'_j,&(DD2)\cr
\endeq
where Ad$(q^h)x=q^hxq^{-h}$ for $x\in U$.
Note that $\cal D$ acts on $V(\la)$ as a scalar $p^{\Delta_\la}$
while $p^d$ acts on $V_z$ as the difference operator $p^dv(z)=v(pz)$.
Setting
$\Phi_i(z)=z^{\Delta_{\mu}-\Delta_{\la}}\tilde\Phi_i(z)$ ($i=1,2$)
we get
\proclaim Proposition \prop{A3}.
\eq
&\Phi_1(pz)=\sigma({\cal R}'(q^{k+2h^{\vee}}))
({\hbox{\rm id}}\ot q^{-2\ovl\rho})
\sum ({\hbox{\rm id}}\ot a^{-2}(b'_i))\Phi_1(z)a'_i,&(DD5)\cr
&\Phi_2(pz)=(q^{-2\ovl\rho}\ot {\hbox{\rm id}})
\sum ({\hbox{\rm Ad}}(q^{kd})a^{-2}(b'_i)\ot
{\hbox{\rm id}}){\cal R}'\Phi_2(z)a'_i.&(DD6)\cr
\endeq

\subsec(A1.2|Two point functions)
Let $\Psi(z_1,z_2)$ be one of the following two point correlation functions:
\eq
&\bra{\nu}\Phi^{\nu V_2}_\mu(z_2)\Phi^{\mu V_1}_\la(z_1)\ket{\la} &(I)\cr
&\bra{\nu}\Phi^{V_2 \nu}_\mu(z_2)\Phi^{\mu V_1}_\la(z_1)\ket{\la} &(II)\cr
&\bra{\nu}\Phi^{\nu V_2}_\mu(z_2)\Phi^{V_1 \mu}_\la(z_1)\ket{\la} &(III)\cr
&\bra{\nu}\Phi^{V_2 \nu}_\mu(z_2)\Phi^{V_1 \mu}_\la(z_1)\ket{\la} &(IV)\cr
\endeq
Here we use the convention that the space suffixed by $1$ (resp. $2$)
always comes in the first (resp. second) component of the tensor product.
For instance, \refeq{I} is the expectation value of the composition
\eq
V(\la)~\goto{\Phi^{\mu V_1}_\la(z_1)}~V(\mu)\ot V_{z_1}
&~\goto{\Phi^{\nu V_2}_\mu(z_2)\ot 1}~V(\nu)\ot V_{z_2}\ot V_{z_1}
{}~\goto{1\ot P}~V(\nu)\ot V_{z_1}\ot V_{z_2} \cr
\endeq
while \refeq{III} corresponds to
\eq
V(\la)~\goto{\Phi^{V_1 \mu}_\la(z_1)}~V_{z_1}\ot V(\mu)
&~\goto{1\ot \Phi^{\nu V_2}_\mu(z_2)}~V_{z_1}\ot V(\nu)\ot V_{z_2}. \cr
\endeq

We shall derive the $q$-KZ equation for the two point functions.
Given finite dimensional representations $\pi_{V}:U' \goto{} \End(V)$,
$\pi_{W}:U' \goto{} \End(W)$,
we set
\eq
R_{VW}^+(z)=&\pi_{V\ot W}({\cal R}'(z)). &(univR)
\endeq

\proclaim Proposition \prop{2ptKZ}. We have
\eq
\Psi(pz_1,z_2)=A(z_1/z_2)\Psi(z_1,z_2), \quad
&\Psi(pz_1,pz_2)=\bigl(q^{-\phi}\ot q^{-\phi}\bigr)\Psi(z_1,z_2), &(2qKZ)\cr
&\phi=\bar{\la}+\bar{\nu}+2\bar{\rho}, \cr
\endeq
where
\eq
A(z)=& R^+(pz)\bigl(q^{-\phi}\ot 1\bigr) \qbox{ for \refeq{I},}&(a)\cr
    =& \bigl(q^{-\bar{\nu}}\ot 1\bigr)R^+(pq^{-k}z)
	\bigl(q^{-\phi+\bar{\nu}}\ot 1\bigr) \qbox{ for \refeq{II},}
&(b) \cr
    =& \bigl(q^{-\phi+\bar{\nu}} \ot 1\bigr)R^+(q^kz)
	\bigl(q^{-\bar{\nu}}\ot 1\bigr) \qbox{ for \refeq{III},} &(c)\cr
    =& \bigl(q^{-\phi}\ot 1\bigr)R^+(z) \qbox{ for \refeq{IV},} &(d)\cr
\endeq
and $R^+(z)=R^+_{V_1V_2}(z)$ is given in \refeq{univR}.
In particular we can write
\eq
&\Psi(z_1,z_2)=z_1^{\Delta_\mu-\Delta_\la}z_2^{\Delta_\nu-\Delta_\mu}
\widetilde{\Psi}(z_1/z_2), \quad
\widetilde{\Psi}(z) \in V_1\ot V_2 \ot F[[z]]. \cr
\endeq
\par
\Proof
We prove the case \refeq{III}. The other cases are similar.

{}From \refeq{DD6} and \refeq{Univ'} we have
\eq
&\fun(\nu,\nu V_2,\mu,z_2,V_1\mu,pz_1)\cr
&=\ffun(\nu,\nu V_2,\mu,z_2,{(q^{-2\ovl\rho}\ot{\hbox{\rm id}})
\sum({\hbox{\rm Ad}}(q^{kd})a^{-2}(b'_i)\ot{\hbox{\rm id}}){\cal R}'},
V_1\mu,z_1,a'_i)\cr
&=\sum\pi_{V_1}(q^{-2\ovl\rho}{\hbox{\rm Ad}}(q^{kd})a^{-2}(b'_i)a'_j)
\ffun(\nu,\nu V_2,\mu,z_2,b'_j,V_1\mu,z_1,a'_i)\cr
&=\sum\pi_{V_1}(q^{-2\ovl\rho}{\hbox{\rm Ad}}(q^{kd})a^{-2}(b'_i)a'_j)
\ot\pi_{V_2}(b'_{j(2)})
\fffun(\nu,b'_{j(1)},\nu V_2,\mu,z_2,V_1\mu,z_1,a'_i)\cr
&=\sum\pi_{V_1}(q^{-2\ovl\rho-\ovl\la}a'_j)
\ot\pi_{V_2}(b'_{j(2)})
\fffun(\nu,b'_{j(1)},\nu V_2,\mu,z_2,V_1\mu,z_1,a'_i)\cr
&=(q^{-2\ovl\rho-\ovl\la}\ot{\hbox{\rm id}})\pi_{V_1\ot V_2}
({\cal R}'(q^k))(q^{-\ovl\nu}\ot{\hbox{\rm id}})
\fun(\nu,\nu V_2,\mu,z_2,V_1\mu,z_1).\cr
\endeq
\qed

Similar argument leads to the $q$-KZ equation for general
$n$ point correlation functions, see \refto{FR}.

The following observations are sometimes useful.
We owe the second one to M. Okado.
\proclaim Proposition \prop{reduc}.
\item{(i)} Suppose $\la$, $\mu$, $\nu$ are dominant integral,
and consider $\Psi(z_1,z_2)$ in the case \refeq{I}.
Then for any $i$ we have
\eq
&\pi_{V_{z_1}}\ot\pi_{V_{z_2}}\bigl(\Delta'(e_i)^{\br{h_i,\nu}+1}\bigr)
\Psi(z_1,z_2)=0,\quad
\wt \Psi(z_1,z_2)= \bar{\la}-\bar{\nu},
\endeq
where $\Delta'=\sigma\circ\Delta$.
\item{(ii)}
Let $\Psi(z_1,z_2)$ be a solution for \refeq{a}. Then the following give
solutions for the other cases:
\eq
&\bigl(q^{-\bar{\nu}}\ot 1\bigr)\Psi(q^{-k}z_1,z_2)\qbox{ for \refeq{b},} \cr
&\bigl(q^{-\phi+\bar{\nu}}\ot 1\bigr)
	\Psi(p^{-1}q^{k}z_1,z_2)\qbox{ for \refeq{c},} \cr
&\bigl(q^{-\phi}\ot 1\bigr)\Psi(p^{-1}z_1,z_2)\qbox{ for \refeq{d}.} \cr
\endeq
\par

\Proof
Let $\Phi=\Phi_\mu^{\nu V_1}(z_1)\circ \Phi_\la^{\mu V_2}(z_2)$.
Then this is an intertwiner $V(\la)\longrightarrow V(\nu)\ot W$
where $W=V_{z_1}\ot V_{z_2}$.
Clearly $\wt \bra{\nu}\Phi\ket{\la}=\bar{\la}-\bar{\nu}$.
Since $V^r(\nu)$ is integrable we have
$\bra{\nu}e_i^{\br{h_i,\nu}+1}=0$.
Setting $l=\br{h_i,\nu}+1$, $x=e_i^l$ and
using \refeq{Hopf2}  we get
\eq
0&=\bra{\nu}x\Phi\ket{\la}
=\sum \pi_W\bigl(a^{-1}(x_{(2)})\bigr)
\bra{\nu}\Phi x_{(1)}\ket{\la}  \cr
&=\pi_{W}\bigl(a^{-1}(e_i^l)\bigr)\bra{\nu}\Phi t_i^l\ket{\la}.
\endeq
Here we have used $e_i\ket{\la}=0$.
This implies
$\pi_{W}(e_i^l)\bra{\nu}\Phi \ket{\la}=0$.
Multiplying the transposition $P$ from the left and
interchanging $z_1$ and $z_2$ we arrive at the assertion (i).
Assertion (ii) can be verified directly.
\qed

\beginsection Appendix 2. Basic hypergeometric series

We recall several facts on the basic hypergeometric series
(see e.g. \refto{GR}).

The basic hypergeometric series is defined by
$$
\hg(a,b;c;z)=\sum_{n=0}^{\infty}{{(a;q)_n(b;q)_n}
\over{(c;q)_n(q;q)_n}}z^n,
\eqno(hg1)
$$
where
$$
(a;q)_n=\prod_{j=0}^{n-1}(1-aq^j), \quad (a;q)_0=1.
$$
The series \refeq{hg1} converges absolutely for $|z|<1$ when $|q|<1$ and can be
analytically continued to the domain $|{\hbox{\rm arg}}(-z)|<\pi$.
We shall also use the $q$-Gamma function and the theta function
\eq
\Gamma_q(z)&={{(q;q)_{\infty}}\over{(q^z;q)_{\infty}}}(1-q)^{1-z},\qquad
\Theta_q(z)=(z;q)_{\infty}(q z^{-1};q)_{\infty}(q;q)_{\infty}. &(Eq8)
\endeq
They enjoy the properties
\eq
&\Gamma_q(z+1)={1-q^z \over 1-q}\Gamma_q(z),\quad \Gamma_q(1)=1,
&(Eq7)\cr
&\Theta_q(z^{-1})=\Theta_q(qz)=-z^{-1}\Theta_q(z), \quad
\Gamma_q(z)\Gamma_q(1-z)={(q;q)_\infty^3(1-q) \over \Theta_q(q^z)}.
\endeq

Consider the difference equation
$$
(q^c-q^{a+b+1})u(q^2z)-\{(q+q^c)+(q^a+q^b)qz\}u(qz)+q(1-z)u(z)=0.\eqno(Eq1)
$$
If we supplement \refeq{Eq1} with the condition that
$u(z)$ has an expansion of the form
$u(z)=z^{\al}\sum_{n=0}^{\infty}a_nz^n$ around $z=0$,
then the solution space of \refeq{Eq1} is two-dimensional over $\C$,
and we can take
$$
\hg(q^a,q^b;q^c;z)\qbox{and}\q z^{1-c}\hg(q^{a-c+1},q^{b-c+1};q^{2-c};z)
$$
as a basis.

We list some useful fomulas below.
\eq
&\hg(q^a,q^c;q^c;z)=\sum_{n=0}^{\infty}{{(q^a;q)_n}\over{(q;q)_n}}z^n
={{(q^az;q)_{\infty}}\over{(z;q)_{\infty}}},&(Eq100)\cr
&\lim_{q\rightarrow 1}{{(q^az;q)_{\infty}}\over{(z;q)_{\infty}}}
=(1-z)^{-a},\cr
&\hg(q^a,q^b;q^c;z)={{(q^{a+b-c}z;q)_{\infty}}\over{(z;q)_{\infty}}}
\hg(q^{c-a},q^{c-b};q^c;q^{a+b-c}z),&(Eq5)\cr
&\hg(q^a,q^b,;q^c;q^{a+b-c})={{(q^{c-a};q)_{\infty}(q;q)_{\infty}}
\over{(q^c;q)_{\infty}(q^{c-a-b};q)_{\infty}}},&(Eq6)\cr
&\hg(q^a,q^b;q^c;z)
={{\Gamma_q(c)}\over{\Gamma_q(b)\Gamma_q(c-b)}}\int_0^1t^{b-1}
{{(tzq^a;q)_{\infty}(tq;q)_{\infty}}
\over{(tz;q)_{\infty}(tq^{c-b};q)_{\infty}}}d_qt.
\endeq
The last formula involves the Jackson integtal defined by
$$
\int_0^1f(t)d_qt=(1-q)\sum_{n=0}^{\infty}f(q^n)q^n.
$$
The following connection formula holds
\eq
\hg(q^a,q^b;q^c;{1\over z})
&={\conn(c,b-a,q^{1-a}z) \over \conn(b,c-a,qz) }
\hg(q^a,q^{a-c+1};q^{a-b+1};q^{c-a-b+1}z)\cr
&+{\conn(c,a-b,q^{1-b}z) \over \conn(a,c-b,qz) }
\hg(q^b,q^{b-c+1};q^{b-a+1};q^{c-a-b+1}z), \cr
&&(Eq4) \cr
\endeq
provided that $|{\hbox{\rm arg}}(-z)|<\pi$, $c$ and $a-b$ are non-integers,
and $z\ne 0$.
%
\catcode`@=11
\newif\ifs@p
\def\refjl#1#2#3#4%
  {#1\def\l@st{#1}\ifx\l@st\empty\s@pfalse\else\s@ptrue\fi%
   \def\l@st{#2}\ifx\l@st\empty\else%
   \ifs@p, \fi{\frenchspacing\sl#2}\s@ptrue\fi%
   \def\l@st{#3}\ifx\l@st\empty\else\ifs@p, \fi{\bf#3}\s@ptrue\fi%
   \def\l@st{#4}\ifx\l@st\empty\else\ifs@p, \fi#4\s@ptrue\fi%
   \ifs@p.\fi\hfill\penalty-9000}
\def\refbk#1#2#3%
  {#1\def\l@st{#1}\ifx\l@st\empty\s@pfalse\else\s@ptrue\fi%
   \def\l@st{#2}\ifx\l@st\empty\else%
   \ifs@p, \fi{\frenchspacing\sl#2}\s@ptrue\fi%
   \def\l@st{#3}\ifx\l@st\empty\else\ifs@p, \fi#3\s@ptrue\fi%
   \ifs@p.\fi\hfill\penalty-9000}
\catcode`@=12
%
%

\def\CMP{Commun. Math. Phys.}
\def\Duke{Duke Math. J.}

\def\IJMPA{Int. J. Mod. Phys. A}

\def\JPA{J. Phys. A: Math. Gen.}
\def\JSP{J. Stat. Phys.}

\def\NP{Nucl. Phys.}
\def\NPB{Nucl. Phys. B}

\def\PL{Phys. Lett.}

\def\RIMS{RIMS preprint}

\par
\bigskip\noindent{\bf References}\medskip
\par
\refis{Bax} \refjl
{Baxter R J,
Spontaneous staggered polarization of the $F$ model}
{\JSP}{9}{(1973) 145--182}
\par

\refis{DFJMN} \refjl
{Davies B, Foda O, Jimbo M, Miwa T and Nakayashiki A,
Diagonalization of the XXZ Hamiltonian by vertex operators}
{\RIMS}{873}{(1992)}
\par

\refis{JMMN} \refjl
{Jimbo M, Miki K, Miwa T and Nakayashiki A,
Correlation functions of the XXZ model for $\Delta<-1$}
{\RIMS}{877}{(1992)}
\par

\refis{FM} \refjl
{Foda O and Miwa T,
Corner transfer matrices and quantum affine algebras}
{\IJMPA}{7}{supplement 1A (1992) 279--302}
\par

\refis{FrJ} \refjl
{Frenkel I B and Jing N H,
Vertex representations of quantum affine algebras}
{Proc. Nat'l. Acad. Sci. USA}{85}{(1988) 9373--9377}
\par

\refis{FR} \refjl
{Frenkel I B and Reshetikhin N Yu,
Quantum affine algebras and holonomic difference equations}
{\CMP}{149}{(1992) 1--60}
\par

\refis{Baxbk} \refbk
{Baxter R J}
{Exactly solved models in statistical mechanics}
{Academic Press, London 1982}

\refis{DJO} \refjl
{Date E, Jimbo M and Okado M,
Crystal base and $q$ vertex operators}
{Osaka Univ. Math. Sci. preprint}{1}{(1991)}
\par

\refis{CP} \refjl
{Chari V and Pressley A,
Quantum affine algebras}
{\CMP}{142}{(1991) 261--283}
\par

\refis{Dri} \refjl
{Drinfeld V G,
A new realization of Yangians and quantum affine algebras}
{Soviet Math. Doklady}{36}{(1988) 212--216}
\par

\refis{(KMN)^2} \refjl
{Kang S-J, Kashiwara M, Misra K, Miwa T, Nakashima T and Nakayashiki A,
Affine crystals and vertex models}
{\IJMPA}{7}{Suppl. 1A (1992) 449--484}
\par

\refis{Reshet} \refjl
{Reshetikhin N Yu,
$S$-matrices in integrable models of isotropic magnetic chains:I}
{\JPA}{24}{(1991) 3299-3309}
\par

\refis{Sogo} \refjl
{Sogo K,
Ground state and low-lying excitations in the Heisenberg
XXZ chain of arbitrary spins}
{\PL}{104A}{(1984) 51-54}
\par

\refis{Kac} \refjl
{Kac V G, Infinite dimensional Lie algebras}
{3rd edition ,Cambridge Univ. Press}{}{(1990)}
\par

\refis{D} \refjl
{Drinfeld V G, Quantum Groups}{Proceedings of ICM Barkeley}
{}{(1986) 798-820}
\par

\refis{Ta} \refjl
{Tanisaki T, Killing forms, Harish-Chandra isomorphisms,
and universal $R$-matrices for quantum algebras}
{\IJMPA}{7}{Suppl. 1B (1992) 941--961}
\par
\par

\refis{D2} \refjl
{Drinfeld V G, On almost co-commutative Hopf algebras}
{Leningrad Math. J.}{1}{(1990) 321-431}
\par

\refis{GR} \refjl
{Gasper G and Rahman M, Basic hypergeometric series}{Cambridge Univ. Press}{}
{(1990)}
\par

\refis{CP} \refjl
{Chari V and Pressley A,
Quantum affine algebras}
{\CMP}{142}{(1991) 261--283}
\par

\refis{Dri} \refjl
{Drinfeld V G,
A new realization of Yangians and quantum affine algebras}
{Soviet Math. Doklady}{36}{(1988) 212--216}
\par

\refis{Ka} \refjl
{Kashiwara M,
On crystal bases of the $q$-analogue of universal enveloping algebras}
{\Duke}{63}{(1991) 465--516}

\par
\refis{Ka2} \refjl
{Kashiwara M,
Global crystal bases of quantum groups}
{\RIMS}{756}{(1991)}
\par

\refis{S} \refjl
{Kulish P P and Reshetikhin N Yu,
Quantum linear problem for the Sine-Gordon equation and higher representations}
{Zapiski nauch. Sem. LOMI}{101}{(1980) 101--110}
\par

\refis{Smbk} \refbk
{Smirnov F A,
Form factors in completely integrable models of quantum field theory}
{Advanced Series in Mathematical Physics {\bf 14}}
{World Scientific, Singapore 1992}

\listreferences
\par

\refis{DJMO} \refjl
{Date, E., Jimbo, M., Kuniba, A., Miwa, T. and Okado, M,
One dimensional configuration sums in vertex models and affine Lie
algebra characters}
{Lett. Math. Phys.} {17} {(1989) 69--77}

\refis{DJKMO} \refjl
{Date E, Jimbo M, Kuniba A, Miwa T and Okado M,
Exactly solvable SOS models:
Local height probabilities and theta function identities}
{\NPB}{290}{[FS20] (1987) 231--273}

\def\uq{U_q\bigl(\widehat{\goth{sl}}\hskip2pt(n)\bigr)}

\refis{MM} \refjl
{Misra K C and Miwa T,
Crystal base for the basic representation of $\uq$}
{\CMP}{134}{(1990) 79--88}

\refis{JMMO} \refjl
{Jimbo M, Misra K C, Miwa T and Okado M,
Combinatorics of representations of $\uq$ at $q=0$}
{\CMP}{136}{(1991) 543--566}

\refis{ABF} \refjl
{Andrews G E, Baxter R J and Forrester P J,
Eight-vertex $SOS$ model and generalized Rogers-Ramanujan-type identities}
{\JSP}{35}{(1984) 193-266}

\refis{ABL} \refjl
{Ahn C, Bernard D and LeClair A,
Fractional supersymmetries in perturbed coset CFT's
and integrable soliton theory}
{\NP}{B340}{(1990) 721}

\refis{FLC} \refjl
{Felder G and LeClair A,
Restricted quantum affine symmetry of perturbed
minimal conformal models}
{\IJMPA}{7}{Suppl. 1A (1992) 239--278}

\refis{Sm} \refjl
{Smirnov F A,
Dynamical symmetries of massive integrable models}
{\RIMS}{772, 838}{(1992)}

\refis{Pas} \refjl
{Pasquier V,
Etiology of IRF models}
{\CMP}{118}{(1988) 335-364}

\refis{Ji} \refjl
{see e.g. Jimbo M, Introduction to the Yang-Baxter equation}
{\IJMPA}{4}{(1989) 3759-3777}

\refis{BS} \refjl
{Bazhanov V V and Stroganov Yu G,
Chiral Potts model as a descendant of the six vertex models}
{\JSP}{51}{(1990) 799--817}

\refis{cl} \refjl
{Clavelli L and Shapiro J A,
Pomeron Factorization in General Dual Models}
{\NPB}{57}{(1973) 490--535}
\par

\refis{Zinv} \refjl
{Baxter R J,
Solvable eight-vertex model on an arbitrary planar lattice}
{Phil. Trans. Roy. Soc.}{A289}{(1978) 315--346}
\par

\refis{AYP} \refjl
{Au-Yang H and Perk J H H,
Critical Correlations in a $Z$-invariant Inhomogeneous Ising Model}
{Physica} {144A} {(1987) 44--104}
\par

\refis{Jg} \refbk
{Jing N H,
On a trace of $q$ analog vertex operators}
{in Quantum Groups, Spring Workshop on Quantum Groups
Argonne National Laboratory 16 April--11 May 1990,
eds. T. Curtright, D. Fairlie and C. Zachos}
{}{World Scientific, Singapore 1992}
\par

\end